\documentclass[%
 aip,
 amsmath,amssymb,
 preprint,%
]{revtex4-1}

\usepackage{graphicx}
\usepackage{dcolumn}
\usepackage{bm}

\usepackage[utf8]{inputenc}
\usepackage[T1]{fontenc}
\usepackage{mathptmx}
\usepackage{amssymb}
\usepackage{mathrsfs}  
\usepackage{etoolbox}

\makeatletter
\def\@email#1#2{%
 \endgroup
 \patchcmd{\titleblock@produce}
  {\frontmatter@RRAPformat}
  {\frontmatter@RRAPformat{\produce@RRAP{*#1\href{mailto:#2}{#2}}}\frontmatter@RRAPformat}
  {}{}
}%
\makeatother
\begin{document}


\title[Analysis of axial waves in visco-elastic complex structural-acoustic systems]{Analysis of axial waves in visco-elastic complex structural-acoustic systems:\\Theory
and experiment}
\author{J. A. Rojas}
\affiliation{Instituto de Investigaci\'on en Ciencias B\'asicas y Aplicadas, Universidad Aut\'onoma del Estado de Morelos, Av. Universidad 1001, 62209 Cuernavaca, Morelos, Mexico.}
\affiliation{Instituto de Ciencias F\'isicas, Universidad Nacional Aut\'onoma de M\'exico, Av. Universidad 1001, 62210 Cuernavaca, Morelos, Mexico.}%
\author{A. Morales}%
\author{L. Guti\'errez}%
 \email{monsi@fisica.unam.mx}

\affiliation{Instituto de Ciencias F\'isicas, Universidad Nacional Aut\'onoma de M\'exico, Av. Universidad 1001, 62210 Cuernavaca, Morelos, Mexico.}%

\author{J. A. Otero}%
\affiliation{Escuela de Ingenier\'ia y Ciencias, Tecnol\'ogico de Monterrey,
Carr. al Lago de Guadalupe Km. 3.5, 52926 Estado de M\'exico, Mexico.}
\author{E. A. Carrillo}
\author{G. Monsivais}
\author{J. Flores} 
\thanks{Deceased}
\affiliation{%
Instituto de F\'isica, Universidad Nacional Aut\'onoma de M\'exico - P.O. Box 20-364, 01000 Ciudad de México, Mexico.
}%

\date{\today}

\begin{abstract}
An experimental and theoretical study of the spectral response of coupled visco-elastic bars
subject to axial oscillations is done. Novel closed formulas for the envelope function and
their width is derived. These formulas explicitly show the role played by energy dissipation.
They show that the internal friction does not affect the width of the envelope of the
individual resonances. The formulation is based on the equations of classical mechanics
combined with Voigt’s viscoelastic model. The systems studied consist of a sequence
of one, two, or three coupled bars, with their central axes collinear. One of the bars is assumed
to be much longer than the others. We discuss the connection between our results
with the concept of the strength function phenomenon discovered for the first time in nuclear
physics. Our formulation is an alternative and exact approach to the approximated
studies based on the fuzzy structure theory that has been used by other authors to describe this
type of couplings. The analytical expressions describe the measurements in the laboratory
very well.
\end{abstract}

\maketitle

\section{\label{sec:1} INTRODUCTION}

It is well known that when a solid elastic bar is excited with an axial harmonic force, a set of resonances whose frequencies are equidistant is produced. The intensities and amplitudes of these resonances depend on the energy dissipation present in the bar. However, a lesser-known phenomenon occurs when the bar has a notch near one of its ends so that the resulting system is a sequence of two coupled bars, one small and one long. The perturbation produced by this modification drastically changes the intensity distribution of the resonances, giving rise to a profile of such distribution that resembles a Lorentzian function very wide, centered at a frequency close to the resonant frequency of the small bar. What makes this phenomenon particularly interesting is that it occurs in a large number of physical systems: mechanical, classical and quantum electromagnetic, nuclear, etcetera, where a structure with a very dense eigenspectrum (called a sea of states) is coupled to a system with a low density of eigenvalues, which causes the so-called strength function phenomenon and giant resonances to arise. In a previous experimental and numerical study, Ref. \cite{28}, it was determined the dependence of giant resonances on the different parameters of a cylindrical aluminum bar with a notch. In the present work, we derive analytical expressions that confirm these results. Closed formulas are derived, which give the parameters of the giant resonance in terms of energy loss due to internal friction and coupling with the atmosphere.

Specifically, the first objective of this paper is to report the results of experimental measurements and derive 
analytical expressions for the envelope function of the spectral response of coupled bar systems subject to axial oscillations.
This kind of enveloping function has been discussed little in the literature. Indeed, while the shape of the curve of a single resonance is a widely discussed topic, the shape of the envelope of a family of resonances is not a sufficiently discussed topic.
It is important to study these envelopes
because that is what an observer detects when using instruments with an insufficient resolution to see the details of the phenomenon studied.
He will not be able to see the individual resonances but only one very broad resonance.
These envelopes are also important when the only thing that matters is knowing the coarse response of the studied system.

On the one hand, in the case of a single resonance, is usually found that on sufficiently simple systems and, as the energy dissipation tends to zero, the form of the square of the resonance curve has the standard form of a Lorentzian. 
On the other hand, in the case of the envelopes, the few works that have studied them have provided rather incomplete information.
Among the few works that study these envelopes is that of reference \cite{15}, and those references mentioned there. These works study how the states of systems governed by quantum mechanics having a special state (the doorway state) are distributed.  They prove that when the density of the energy spectrum is equal to a constant the distribution is a Breit-Wigner function, which essentially is the same as a Lorentzian function. However, in general the energy spectra of the systems do not meet this hypothesis, and therefore, that prediction may be wrong. Moreover, these discussions do not provide information about the values that the parameters that characterize the Lorentzian should have. 

Additionally, as mentioned above, those studies were only in the context of the quantum mechanics.  So, the subject, for elastic systems within the frame of classical mechanics has not been studied. This is doing in the present paper. We will study systems where the laws of the elasticity are the ones that governs their movement. The systems are similar to those studied in references \cite{27,28} and in references \cite{nunes2006,langley1999}. They consist of one, two or there coupled bars subjected to axial vibrations. 
The structures were designed so that one of the bars forming the composite system has a very high spectral density when considered isolated from the other bars. The other bars have a very low density.
We have experimentally observed that in the coupled bars the spectral density is not equal to a constant. 
Therefore, the envelope of the single resonances should not be expected to be a perfect Lorentzian. 

The second objective of this article is to discuss the connection between two quite different point of view with which the systems considered here can be studied.
One of them is based on approximated approaches. Among them are for example: statistical energy analysis (SEA) \cite{lyon1995},
spectral element method (SEM) \cite{nunes2006}, 
fuzzy structure theory (FST) \cite{soize1993,pierce1995,strasberg1996}, Belyaev smooth function approach (BSFA) \cite{belyaev1986,belyaev1992,belyaev1993},  
hybrid method vibration Analysis (HMVA) \cite{langley1999}, etc. Several of them are based on the FST, in which it is considered that the system consists of a master structure that has a set of fuzzy couplings connected to it. Usually, the master structure has a small spectral density compared to the spectral density of the fuzzy couplings. In this context the underlying hypothesis is that the details of the fuzzy couplings are not essential for the overall or macroscopic description of the system.

The other point of view is an exact approach in which phenomena appear analogous to
those discovered in the 1940's and that was later found to be present 
in other systems as well, both quantum \cite {10,11,12,13,14,15,16,17,18,19,20,21,22,23,24} and classical
\cite{25,26,27,28,29}. These are: the strength function phenomenon, doorway states and giant resonances.
This approach constitutes an exact and alternative description to the approximate studies based on the fuzzy structure theory mentioned above. 
Unfortunately, 
very little has been discussed in the literature about the connection between these two points of view. 
As in the case of the fuzzy structure theory, the doorway states appear when two or more systems, which have very different spectral density, are brought into interaction. One with a low and usually simple states density, while the others with a high and usually more complicated density of states forming a `` sea '' of states. As these systems interact, the states of the system with a low-density spectrum act as doorway states to the coupled system. When a doorway is excited then, as time goes by, the injected energy is distributed very efficiently among the eigenstates of the composite system whose eigenvalue is close to that of the doorway. In addition, the amplitude with which each of them are excited has a quasi-Lorentzian envelope, giving rise to a giant resonance. 

	This article discusses the connection between these two points of view. Experimental results are presented  and two analytical formulations are discussed: one of them an exact formulation and the other an approximated formulation. These mathematical expressions explicitly show that the details of the attached couplings have little influence on the gross response of the coupled system which confirms the hypothesis assumed in the fuzzy structure theory. 

In section \ref{sec:2}, we discuss the structure of the frequency spectrum of 
the system show in Fig.~1. Section \ref{sec:3} deals with the derivation of analytical expressions for the different resonances studied here including closed formulas for the width $({\rm FWHM})$ of the resonance curves. We also present numerical and experimental results. In particular, subsections \ref{sec:3a} and \ref{sec:3b} focus on the giant and common resonances, respectively, that exist in bars with a groove. 
In subsection \ref{sec:3c} it is shown that in bars without a groove there are no giant resonances, so only common resonances are studied.
Finally, in section \ref{sec:7}, are the conclusions.

\section{\label{sec:2}SPECTRAL STRUCTURE OF THE COMPOSED BAR}

In this section we discuss the behaviour of the system shown in Fig.~1 when excited by compressional waves. The system consists of a thin metallic complex bar formed by three circular cylinders of lengths $L, \epsilon,$ and $\ell$ as shown. The radii of the cylinders are, respectively, $r_{L}, r_{\epsilon}=\eta r_{L}$, and $r_{\ell}=r_{L}$, with $\eta$ a real number between 0 and 1.  The two cylinders of ratio $r_{L}$ have lengths that fulfill $L\gg\ell$. Therefore, the cylinder of length $L$ has a higher spectral density than the others. The small cylinder of length $\epsilon$ is called groove, it is shorter and narrower than the others. 

\begin{figure}
	\noindent \begin{centering}
	\includegraphics[scale=0.7]{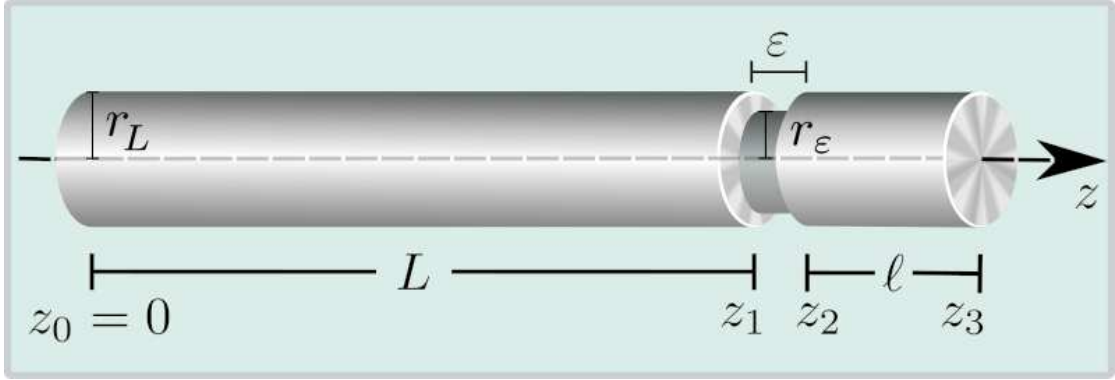}
	\par\end{centering}
	\caption{The bar is formed by three circular cylinders of lengths
	$L,\varepsilon,$ and $\ell$ and radii $r_{L},r_{\varepsilon}=\eta r_{L}$
	and $r_{\ell}=r_{L}$, with $\eta\in\left[0,1\right]$. The values
	used in the experiment are $L=3.607\,\mathrm{m}$,
	$\ell=0.0498\,\mathrm{m}$, $\varepsilon=0.0016\,\mathrm{m}$, $r_{L}=0.25\,\mathrm{in}$,
	$\eta=r_{\varepsilon}/r_{L}=0.236$,
	$v_{\mathrm{c}}=4984\,\mathrm{m/s}$, $E=67.19\,\mathrm{GPa}$, and
	$\rho=E/v_{\mathrm{c}}^{2}$.}
	\end{figure}	
	
The bar was excited at the right extreme $z_3=L+\epsilon+\ell$ by an axial force $F_{\mathrm{ext}}(t)$ of the form

	\begin{equation}
	F_{ext}(t)=h_0\sin (\Omega t)\label{eq:1 fuerza}
 	\end{equation}

\noindent for different values of the frequency $\Omega$. 
The response of the bar was studied by analysing  the amplitude of its oscillations at its left end $z_0=0$.

For convenience, in what follows we will use the name of {\it common resonance} 
(or {\it single resonance})
interchangeably to refer to either the oscillation of the bar itself when an oscillatory force is applied with a frequency equal to one of its natural frequencies, or to refer to the mathematical expression that gives the amplitude of the oscillation as a function of the frequency in a range of frequencies around the natural frequency, or to refer to the plot of this function.
This plot will also be called a resonance curve.
Furthermore, when we have a function for which it is not certain that it is a Lorentzian but whose graphical representation is similar, we will say that it is a quasi-Lorentzian function.

In Fig.~2 are plotted four different curves associated with the response of the bar.
We will first discuss the dashed blue curve. The other three curves will be discussed in the next section. This dashed blue line and the blue lines of Figs.~3 an 4 were calculated using Eq.~(\ref{eq:7 ampl acel}) which was taken from Ref.~\cite{28} after omitting the transient (Eq.~(21) of that reference). Its derivation is briefly reproduced here in the appendix.
The dashed blue curve is indeed   
a sequence of dashed blue vertical curves whose shape is approximately that of a very narrow Lorentzian. 
So, each dashed blue vertical line in Fig.~2 is actually a couple of lines, one going up and the other going down. 
This sequence of cuasi-Lorentzians apparently separated from each other actually form a single continuous line made up of them joined at the bottom. 
As an example, in Figs.~3 and 4 several of these curves were calculated and plotted as blue lines with a very elongated horizontal axis scale to clearly show its shape. 
Each of these curves corresponds
to an axial vibrational resonance of the bar and
it is what we call {\it common} or {\it single} resonance.
The horizontal axis shows the values of the frequency $f$ 
of the exciting force of Eq.~(\ref{eq:1 fuerza}). 
The corresponding values of the square amplitude of the acceleration at $z_0=0$ 
are shown on the vertical axis. 
At this point it is convenient to make the following warning: for simplicity, in our analytical discussion, the frequency will be expressed in radians per second and will be denoted as $\Omega$. On the other hand, in the discussion of our figures the frequency will be expressed in hertz and will be denoted as $f$, with $f=\Omega /2\pi$. This is justified because usually, our figures will display readings taken directly from the devices of the laboratory. 

\begin{figure}
	\includegraphics[width=\textwidth]{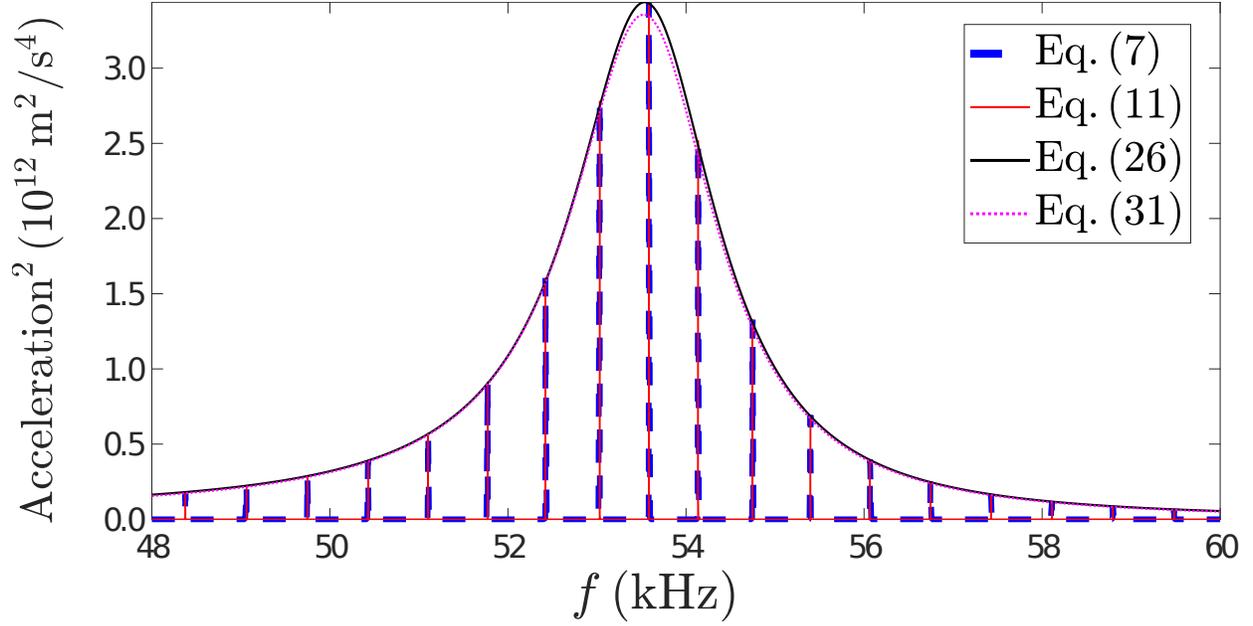}
	\caption{\label{fig:2} The dashed blue line is the plot of the
	square amplitude of the acceleration
	at $z_0=0$ of the bar of Fig.~1 as a function of
	frequency when the bar is excited from the right
	by the force given in Eq.~(\ref{eq:1 fuerza}). It was calculated with the exact 
	formula~(\ref {eq:7    ampl acel}).
    Each dashed blue vertical line represents a common resonance of the system 
    and is drawn in more detail in Fig.~3 or 4.
    The red line is an approximation of the dashed blue line 
    as indicated in the text.
    The black and magenta curves are envelopes of these resonances under different  
    approximations.
    Here $\lambda = 0.12\,{\mathrm{Pa\cdot s}}$ and $h_{0} = 1.0 \, \mathrm {N}$. 
    The values of the 
	other parameters are indicated in Fig.~1,  except that of $\eta $, for which was 
	necessary to use an  effective value $ \eta_{\mathrm {eff}} = 0.16 $.}
	\end{figure}

	\begin{figure*}
	\includegraphics[width=1\textwidth]{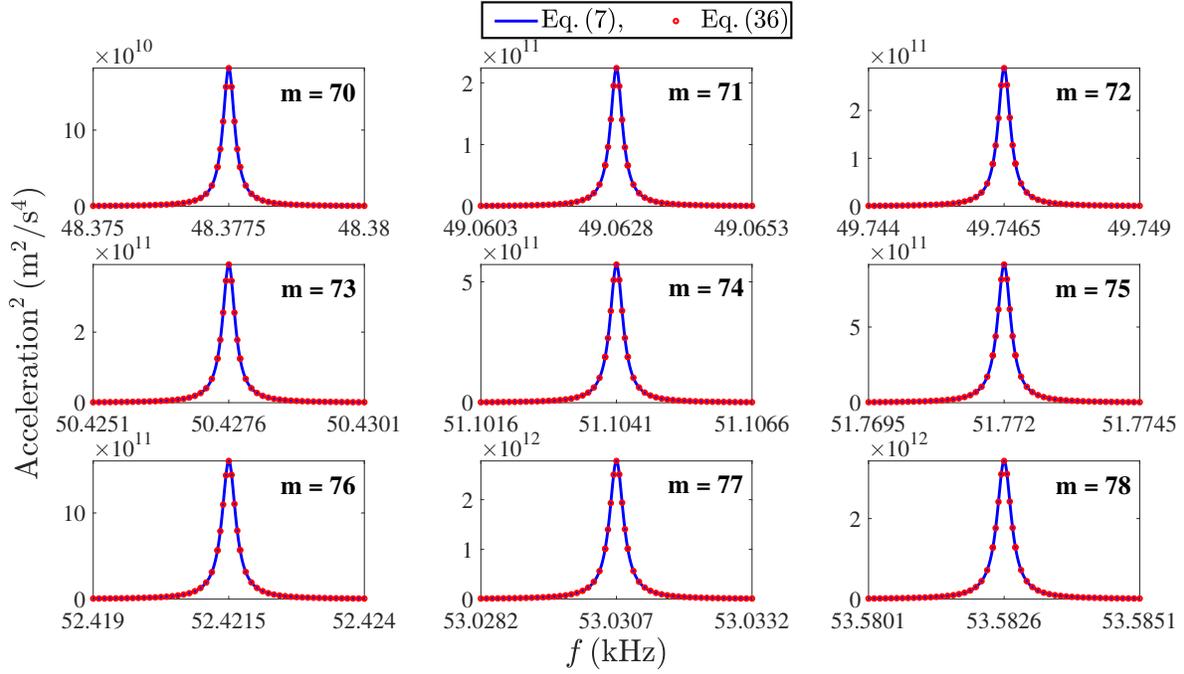}
	\caption{\label{fig:3}
	The blue line of each of these nine graphs corresponds 
	to one of the first nine dashed blue vertical lines in Fig.~2.
    These plots have a very elongated horizontal axis scale in order to clearly show their
    shapes. They were calculated with the exact formula 
      ~(\ref {eq:7 ampl acel}). 
	Each of these resonances was approximated by a quasi-Lorentzian 
	function given by Eq.~(\ref {eq:36}). 
	The red dots represent the plot of such approximation.}
	\end{figure*}

	\begin{figure*}
	\includegraphics[width=1\textwidth]{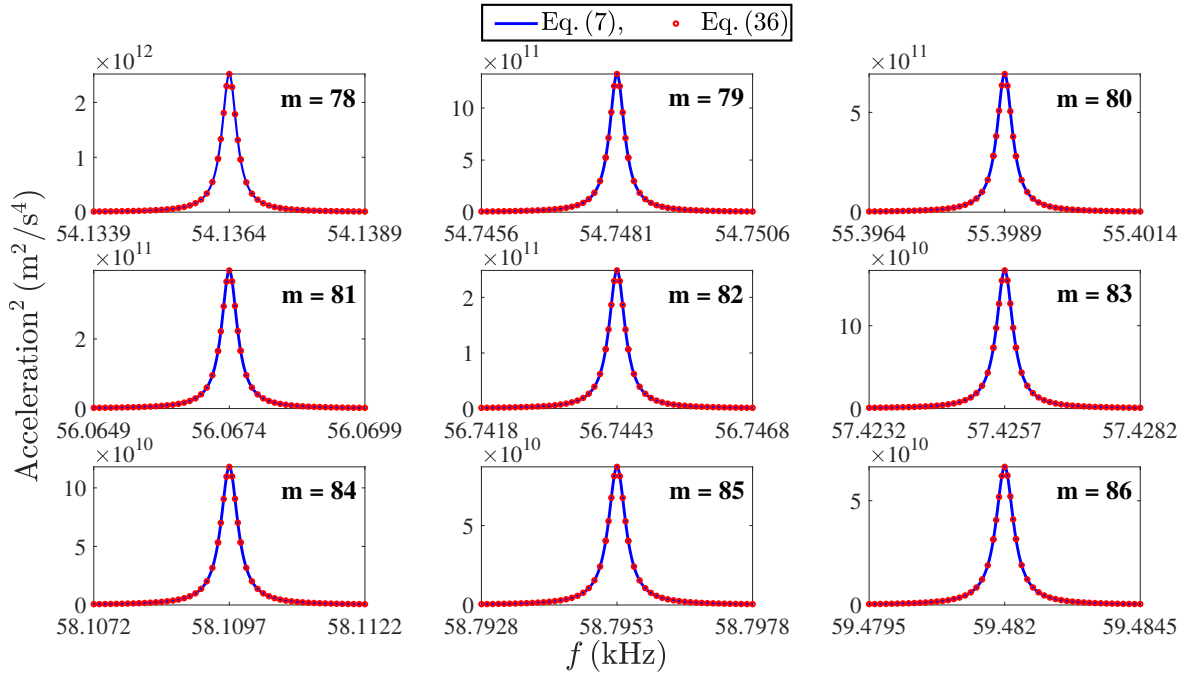}
	\caption{\label{fig:4} Same as in Fig.~3 but for the last nine blue dashed vertical 
	lines of Fig.~2.}
	\end{figure*}

The blue dashed line in Fig.~2 is the theoretical prediction of the experimental results reported in Ref.~\cite{28}, which were plotted in figure 8 of that reference. The blue line reproduces very well several of the observed effects.
The curve is a refinement of the results reported in that reference since higher precision was used here. This allowed the widths of the 18 common resonances shown in Figs.~3 and 4 to be determined with great precision. Their values are plotted in Fig.~8 by means of green stars. It is observed that the theoretical widths are much smaller than the experimental widths. However,
as we will see later, these differences do not affect the envelope curve of the theoretical and experimental common resonances (black curve).

In Fig.~2 it is seen that the intensity of the common resonances is not uniform, being greater for those around the frequency $f_{\mathrm {MAX}}=53.58259~$kHz which is a value near to the first resonance frequency $f_{\ell}=v_{\mathrm{c}}/2\ell=50.0~$kHz of an isolated cylinder equal to the short cylinder of length $\ell$. 
The value of $f_{\mathrm {MAX}}$ depends on the parameter $\eta$ in such a way that when $\eta \to 0$ (that is, when the interaction between the cylinders of length $L$ and $\ell$ tends to zero) the value of $f_{\mathrm {MAX}}$ gets closer and closer to $f_{\ell}$. 
Therefore, the complex bar of Fig.~1 seems to be intrinsically more efficient in absorbing energy for frequencies near the eigenfrequency $f_{\ell}$. 
To construct this plot  an effective value of $\eta$  was used ($\eta_{\mathrm {eff}}=0.160\neq\eta$). The use of an effective $\eta$ is needed to take into account the nesting (or stretching) effect of cylinder 2 on cylinders 1 and 3 when the bar is being compressed (or stretched) during the axial vibrations, see Ref.~\cite{30}. The exact position of the blue quasi-Lorentzian curves is very sensitive to small variations of $\eta_{\mathrm {eff}}$.

The very particular bell-shaped appearance in which the intensities of the resonances are distributed is the so called 
{\it strength function phenomenon} and it occurs when a resonance of the master 
structure is embedded inside the set of resonances of the attached structures as
is the case of the complex bar of Fig.~1.  
In this paper, we will call {\it giant resonance} to the envelope
(as represented by the black curve)
of the set of common resonances when their intensities are distributed as shown in Fig.~2.
A giant resonance is detected as a single very 
wide resonance (the black curve)
when an observer uses instruments with low resolution (See Figs.~6 and 7 in Ref.~\cite{14}).

We see then that the formulation derived from Ref.~\cite{28} predicts, through a numerical calculation, the existence of the strength function phenomenon.
However, in that reference no analytical expressions were derived for it, which would allow a formal prediction of them and establish analytical relationships between the parameters that characterize them. 

Since in Refs.~\cite{27} and \cite{28} there was not analytic expression for the envelope curve it was constructed after numerically determining the set of common resonances.
Once all these resonances were determined, their envelope was built by fitting a function, whose shape was suggested by the intensity and distribution of these resonances, using the least square method. The suggested shape was that of a Lorentzian. Something similar was done to determine the envelope of the set of the experimental resonant curves. It was found that both, the family of calculated resonant curves and the family of the experimental resonant curves, admit the same envelope.
In contrast, in the present paper, the envelope function of Fig.~2 was not obtained by any curve fitting. It was calculated with the formula derived below.

In a homogeneous bar without a groove, the intensities
of the common resonances follow a very different pattern as can be seen in Fig.~5.
The vertical dashed blue lines are again very narrow quasi-Lorentzian curves and correspond to the common resonances of the bar. As can be seen, the intensity of the resonances decreases monotonically, so the strength function phenomenon is not present. Therefore the bars without groove do not have giant resonances. 
The procedure to obtain this plot will be discussed later in connection with the exact expression (\ref{eq:7 ampl acel}) with $\eta=1$. 
Also in this case an expression for the envelope curve 
(represented by a black line in Fig.~5) is obtained. But now the expression is easily obtained, which contrasts with the procedure that must be followed to obtain the giant resonance curve of Fig.~2.

	\begin{figure}
	\includegraphics[width=\textwidth]{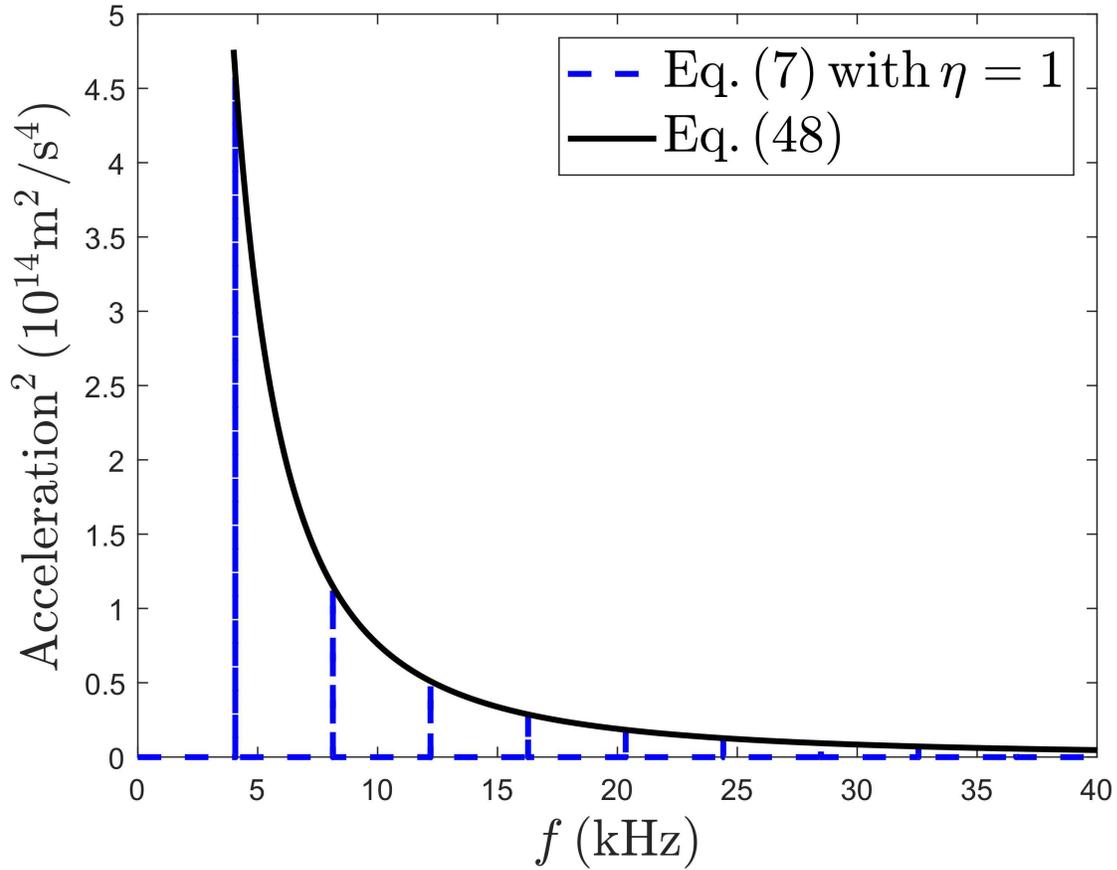}
	\caption{\label{fig:5}The dashed blue line is the square magnitude of acceleration
	calculated at the extreme left of a bar without groove as a function of
	frequency when the bar is excited from the right
	by the force given in Eq.~(\ref{eq:1 fuerza}). 
	Each vertical blue line is a very thin quasi-Lorentzian curve and it
	corresponds to a common resonance
	of the bar.
	The black curve is an envelope of these resonances whose expression is 
	$(\kappa/4\pi^2) f^{-2}$, 
	with $\kappa=\left(2h_{0}	
	v_{\mathrm{c}}^{2}/\left(\pi R^{2}{\mathcal L}\lambda\right)\right)^{2}$, being
	${\mathcal L}=0.5\,\mathrm{m}$ the length of the bar without groove and $R=0.25\,\mathrm{ in}$ its
	radius.}
	\end{figure}

Before presenting the derivation of the formalism, we should make
the following comments. The two types of resonances discussed here have very different characteristics. The common resonances correspond to real oscillations of the elastic body when it is excited by an oscillatory force of frequency equal to one of its normal frequencies. 
These resonances appear in bars both with and without grooves. On the other hand, a giant resonance is the envelope of the family of common resonances of a bar having a groove. Another difference is associated with energy. 
It is well known in the different fields of physics that as the dissipation of energy increases the width of the curve of a common resonance also increases.
Instead, according to the numerical results obtained in ref.~\cite{28},
the width of a giant resonance is not affected by the loss of energy. 
As a consequence, the energy loss cannot be obtained by analyzing the giant resonances only.
Nevertheless, the existence of the giant resonance indicates that energy is absorbed more efficiently by states whose eigenfrequency is near or equal to the frequency of a normal mode of the 
master structure.

\section{\label{sec:3} ANALYTICAL EXPRESSIONS FOR THE CURVES ASSOCIA-~TED WITH THE DIFFERENT RESONANCES}

\subsection{\label{sec:3a} \bf{Giant resonance (envelope curve of the common resonances) for a bar with a groove}}

\noindent 

\noindent The model used in this paper to describe the energy dissipation is the Voigt model for axial waves in elastic bodies Refs.~\cite{31} and \cite{32}. The model introduces a parameter $\lambda$ called {\it coefficient of internal friction} or {\it coefficient of viscosity}.
The following equation is the expression for the acceleration at the left end of the bar when it is excited on the right end by the external force $F_{ext}(t)$ given by  
Eq.~(\ref{eq:1 fuerza}). The expression was taken from Ref.~\cite{28}  
where the transitory part has been eliminated. The procedure to obtain this expression is briefly summarized in the  appendix.

	\begin{equation}
	\frac{d^{2}A\left(0,t\right)}{dt^{2}}=\frac{h_{0}v_{\mathrm{c}}\Omega}{\pi r_{L}^{2}E}\mathrm{Re}\left( 
	\frac{\sqrt{1+i\frac{\lambda\Omega}{E}}e^{i\Omega\,t}}{G}\right),
	\label{eq:2 aceleracion}
	\end{equation}

\noindent 
where 

	\begin{equation}
	G=\sum_{j=1}^4 {\left(-1\right)}^{j+1} \;c_j \;\sinh\left(i\frac{a_j}{\sqrt{1+i\;b}}\right), 
	\;b=\frac{\lambda \Omega }		{E},
	\label{eq:3}
	\end{equation}
	
\vskip10mm

	\[
	a_1 =\frac{\Omega \left(L+l+\varepsilon \right)}{v_\mathrm{c}},\;
	a_2 =\frac{\Omega \left(L+l-\varepsilon \right)}{v_\mathrm{c}},\;
	\]
	
\begin{equation}
	a_3 =\frac{\Omega \left(L-l+\varepsilon \right)}{v_\mathrm{c}},\;
	a_4 =\frac{\Omega \left(L-l-\varepsilon \right)}{v_\mathrm{c}},
\label{eq:4}
\end{equation}

\vskip10mm

	\[
	c_1 =\frac{1}{4}\left(\eta^2 +\frac{1}{\eta^2 }+2\right),\;
	c_2 =\frac{1}{4}\left(\eta^2 +\frac{1}{\eta^2 }-2\right),\;
	\]
	
\begin{equation}
	c_3 =c_4 =\frac{1}{4}\left(\eta^2 -\frac{1}{\eta^2 }\right),
\label{eq:5}
\end{equation}

\noindent here $i^2=-1$. Taking the real part of the right-hand side member
Eq.~(\ref{eq:2 aceleracion}) one obtains:

\begin{equation}
\frac{d^{2}A\left(0,t\right)}{dt^{2}}=
\mathbb {A}\cos\left(\frac{\arctan\left(\frac{\lambda\Omega}{E}\right)}{2}
+\Omega t-\arctan\left(\frac{\mathrm{Im}\,G}{\mathrm{Re}\,G}\right)\right),\label{eq:6}
\end{equation}
where

	\begin{equation}
	\mathbb {A}=\frac
	{   
	\Omega h_{0}v_\mathrm{c}\left(\Omega^{2}\lambda^{2}+E^{2}\right)^{\frac{1}{4}}
	}
	{
	\pi r_L^2 E^{3/2}\sqrt{\left(\mathrm{Re}\,G\right)^{2}+	
                          \left(\mathrm{Im}\,G\right)^{2}}
    },
    \label{eq:7 ampl acel}
   \end{equation}
   
\noindent represents the acceleration amplitude and $\mathrm{Re}\,G$ and
$\mathrm{Im}\,G$  the real and imaginary part of $G$ respectively.

We shall now derive from Eq.~(\ref{eq:7 ampl acel}) the envelope forming  the
giant resonance. This task, as we will see, is neither straightforward nor trivial since the strength function and the giant resonance are not explicitly exhibited in the original formulation. 

The analytical form of $\mathrm{Re}\,G$ and $\mathrm{Im}\,G$ 
functions have a very large 
number of terms. However, taking into account that for the values used in the experiment, 
$E\gg \Omega\lambda \Rightarrow \Omega^2 \lambda^2+E^2\approx E^2$,
$b=\lambda\,\Omega/E\ll 1 \Rightarrow b^n~\approx 0$ for $n \ge 2$,  
etc., it is possible to make a selection of the significant terms and considerably simplify the
expressions. Then, the following approximations are valid:
	\[
	\frac{1}{\sqrt{1+i\;b\;\;}} \approx 1-i\frac{b\,}{2}
	\]
	\begin{equation}
	\Rightarrow\;\sinh\left(i\frac{a_j}{\sqrt{1+i\,b}}\right) \approx -\sinh\left(\frac{a_j\,b}{2}-i\,a_j  \right),  \label{eq:8} \\
	\end{equation}	
	\begin{equation}
	\sinh\left(\frac{a_j \,b}{2}\right) \approx
\frac{a_j \,b}{2},\;\;\cosh\left(\frac{a_j \,b}{2}\right) \approx  1, \label{eq:9}
	\end{equation}

	\begin{equation}
	\sqrt{E^2 +\Omega^2 \,\lambda^2 } = E \sqrt{1 +b^2 }  \approx E\label{eq:10}
	\end{equation}

\noindent and Eq.~(\ref{eq:7 ampl acel}) reduces to
\begin{equation}
\mathbb {A}_{\mathrm{approx}}=\frac{\Omega h_{0}v_\mathrm{c}}{\pi E r_{L}^{2}\sqrt{\left(\mathrm{Re}\,G^{\prime}\right)^{2}+\left(\mathrm{Im}\,G^{\prime}\right)^{2}}},
\label{eq:11 ampl acel aprox}
\end{equation}
where
	\begin{equation}
  \left(\mathrm{Re}\,G^{\prime}\right)^{2} = \sin^{2}\left(\arctan\left(\frac{p_\mathrm{c}}{p_\mathrm{s}}\right)+\frac{\Omega\,L}{v_\mathrm{c}}\right)\,
	\left(p_{\mathrm{c}}^2 + p_\mathrm{s}^2\right) \label{eq:12}\\
	\end{equation}
	\begin{equation}
	\left(\mathrm{Im}\,G^{\prime}\right)^{2} = \sin^{2}\left(\arctan\left(\frac{q_\mathrm{c}}{q_\mathrm{s}}\right)+\frac{\Omega\,L}{v_\mathrm{c}}\right)\,
	\left(q_{\mathrm{c}}^2 + q_\mathrm{s}^2\right)\label{eq:13}\\
	\end{equation}
	
	\begin{eqnarray}
 p_\mathrm{s} &=&~~ \sin\left(\arctan\left(\frac{{p_\mathrm{sc}}}{p_\mathrm{{ss}}}\right)+\frac{\Omega \,l}{v_\mathrm{c}}\right)\, \sqrt{
  p_\mathrm{{sc}}^2 + p_\mathrm{{ss}}^2}\label{eq:14}\\
  p_\mathrm{c} &=& -\sin\left(\arctan\left(\frac{{p_\mathrm{cc}}}{p_\mathrm{{cs}}}\right)+\frac{\Omega \,l}{v_\mathrm{c}}\right)\, \sqrt{
  p_\mathrm{{cc}}^2 + p_\mathrm{{cs}}^2} \label{eq:15}\\
  q_\mathrm{s} &=& ~~\sin\left(\arctan\left(\frac{{q_\mathrm{sc}}}{q_\mathrm{{ss}}}\right)+\frac{\Omega \,l}{v_\mathrm{c}}\right)\, \sqrt{
  q_\mathrm{{sc}}^2 + q_\mathrm{{ss}}^2} \label{eq:16}\\
  q_\mathrm{c} &=& ~~\sin\left(\arctan\left(\frac{{q_\mathrm{cc}}}{q_\mathrm{{cs}}}\right)+\frac{\Omega \,l}{v_\mathrm{c}}\right)\, \sqrt{
  q_\mathrm{{cc}}^2 + q_\mathrm{{cs}}^2 }\label{eq:17}
\end{eqnarray}
and
\begin{eqnarray}
p_{\mathrm{ss}} &=&\frac{b}{2}\;\cos\left(\frac{\Omega \varepsilon }{v_\mathrm{c}}\right)\left(a_1\,c_1-a_2\,c_2-a_3\,c_3+a_4\,c_4 \right)\qquad \label{eq:18}\\
p_{\mathrm{sc}} &=&\frac{b}{2}\;\sin\left(\frac{\Omega \varepsilon }{v_\mathrm{c}}\right)\left(a_1\,c_1+a_2\,c_2+a_3\,c_3+a_4\,c_4 \right) \label{eq:19}\\
p_{\mathrm{cs}} &=&\frac{b}{2}\;\sin\left(\frac{\Omega \varepsilon }{v_\mathrm{c}}\right)\left(a_1\,c_1+a_2\,c_2-a_3\,c_3-a_4\,c_4 \right) \label{eq:20}\\
p_{\mathrm{cc}} &=&\frac{b}{2}\;\cos\left(\frac{\Omega \varepsilon }{v_\mathrm{c}}\right)\left(-a_1\,c_1+a_2\,c_2-a_3\,c_3+a_4\,c_4 \right) \label{eq:21}\\
q_{\mathrm{ss}} &=&\sin\left(\frac{\Omega \varepsilon }{v_\mathrm{c}}\right)\left(-c_1-c_2-c_3+c_4 \right) \label{eq:22}\\
q_{\mathrm{sc}} &=&\cos\left(\frac{\Omega \varepsilon }{v_\mathrm{c}}\right)\left(c_1-c_2-c_3-c_4 \right) \label{eq:23}\\
q_{\mathrm{cs}} &=&\cos\left(\frac{\Omega \varepsilon }{v_\mathrm{c}}\right)\left(c_1+c_2-c_3+c_4 \right) \label{eq:24}\\
q_{\mathrm{cc}} &=&\sin\left(\frac{\Omega \varepsilon }{v_\mathrm{c}}\right)\left(c_1-c_2+c_3+c_4 \right)\label{eq:25}
\end{eqnarray}

\noindent In Fig.~2 are shown the plots of $\mathbb {A}^2$ and 
${\mathbb {A}}_{\mathrm{approx}}^2$ as function of the frequency $f$.
These plots were built using  Eqs.~(\ref{eq:7 ampl acel}) and~(\ref{eq:11 ampl acel aprox})
respectively.
The plot of $\mathbb {A}^2$ is the dashed blue line 
and the plot of ${\mathbb {A}}_{\mathrm{approx}}^2$ is the red line. 
It is clear that the approximate function
is very close to the exact one. 
The figure shows that the same common resonances appear in both graphs. In addition, they are at the same place and with the same intensity.
Note that the coefficient $\lambda$ appears 
as a linear factor in Eqs.~(\ref{eq:18})-(\ref{eq:21}) through the parameter $b$.

In these equations two quotients appear, one $\Omega L/v_{\mathrm{c}}$
much larger than the other $\Omega\ell/v_{\mathrm{c}}$. 
And it is clear that the number of oscillations of the trigonometric functions whose argument contains 
$\Omega L/v_c$ is much larger than when its argument is  $\Omega \ell/v_c$. 
The trigonometric functions with argument  
$\Omega L/v_{\mathrm{c}}$
repeat their value with opposite sign 
every time the frequency increases 
$\pi v_{\mathrm{c}}/L~=~4341~\mathrm{rad/s}$ ~ $=~690.9~\mathrm{Hz}$, which is the
average separation between the vertical lines in Fig.~2. 
This behavior can be explicitly seen  in the function $\mathrm{Im}\,G^{\prime}$ in which
all the arguments of the trigonometric functions contain $\Omega L/v_{\mathrm{c}}$. 
In Fig.~6 it is seen that indeed the variations of $\mathrm{Im}\,G^{\prime}$ 
occur with the same ``frequency'' as the variations of $\mathbb {A}_{\mathrm {approx}}$. But 
one can also see that the zeros of $\mathrm{Im}\,G^{\prime}$ are very 
close to the position of the maxima of $\mathbb {A}_{\mathrm{approx}}$, which fix the points where the envelope
forming the giant resonance must pass. This property is crucial to 
obtain the envelope we are looking for, that is, the envelope of the squared common resonance curves. 
We denote it as~${\mathbb {C}}^{\mathrm{gia}}(\Omega)$.

	\begin{figure}
	\includegraphics[width=\textwidth]{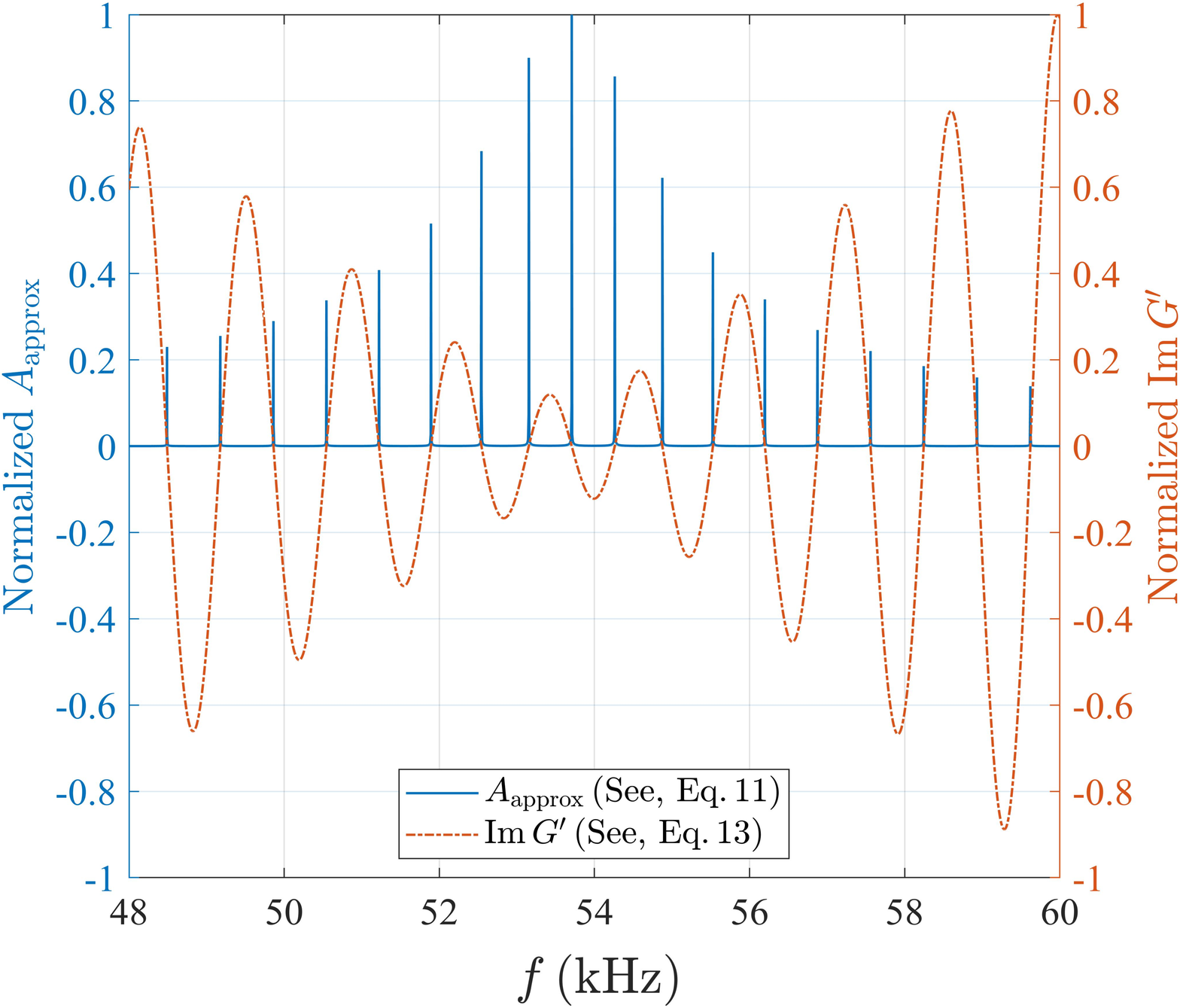}
	\caption{\label{fig:6}
	Normalized plots of ${\mathbb {A}}_{\mathrm{approx}}$ and ${\mathrm {Im}}\,G^{\prime}$ as a function 
	 of the frequency $f$. The maxima of  ${\mathbb {A}}_{\mathrm{approx}}$ are in the zeros of  
	${\mathrm {Im}}\, G^{\prime}$. 
 	}
	\end{figure}

On the other hand, $\mathrm{Re}\,G^{\prime}$ is the product of one
function that oscillates slowly in space with another that has the large spatial frequency.
The first is the modulating function and the second is the modulated function. 
Therefore, to obtain~${\mathbb {C}}^{\mathrm{gia}}(\Omega)$,
the factor $\sin^{2}\left(\arctan\left(\frac{p_{c}}{p_{s}}\right)+\frac{\Omega L}{v_{\mathrm{c}}}\right)$
of  Eq.~(\ref{eq:12}) is set equal to its maximum value  and the resulting function is introduced in
the square of Eq.~(\ref{eq:11 ampl acel aprox}) with $\mathrm{Im}\,G^{\prime}=0$. One obtains

	\begin{equation}
	{\mathbb {C}}^{\mathrm{gia}}\left(\Omega\right)=\left(\frac{\Omega h_{0}v_\mathrm{c}}
	{\pi r_L^{2}\,E}\right)^{2}\frac{1}{p_{s}^{2}+p_{c}^{2}},
	\label{eq:26 res gig}
	\end{equation}
which is one of the expressions we wanted to obtain and is in fact
one of the most important result of this paper. In Fig.~2 an excellent agreement between 
$\mathbb {A}^2$,  $\mathbb {A}_{\mathrm{approx}}^2$
and their envelope ${\mathbb {C}}^{\mathrm{gia}}\left(\Omega\right)$ is seen.

\vskip3mm

\noindent {\bf  Approximation of Eq.~(\ref{eq:26 res gig}) by means of a quasi-lorentzian function}

\vskip3mm

\noindent In order to show that a quasi-Lorentzian function is a good approximation to the function given by 
 Eq.~(\ref{eq:26 res gig}), we use the Taylor expansion for $p_s^2$ and $p_c^2$ of Eqs.~(\ref{eq:14}) and (\ref{eq:15}) around their minima to obtain 

\begin{equation}
p_s^2 \approx \left(\frac{\ell}{v_\mathrm{c}}\right)^2\,P_s^2,\;\;\;p_c^2 \approx \left(\frac{\ell}{v_\mathrm{c}}\right)^2\,P_c^2, 
\label{eq:27}
\end{equation}
where
\begin{equation}
P_s^2 ={{\left[\Omega -\frac{v_\mathrm{c}}{\ell}\,{\left(n\,\pi -\arctan\left(\frac{p_{sc}}{p_{ss}}\right)\right)}\right]}}^2 \,\left(p_\mathrm{{sc}}^2 +p_\mathrm{{ss}}^2\right),\label{eq:28}\\
\end{equation}
\begin{equation}
P_c^2 = {{\left[\Omega -\frac{v_\mathrm{c}}{\ell}\,{\left(n\,\pi -\arctan\left(\frac{p_{cc}}{p_{cs}}\right)\right)}\right]}}^2 \,\left(p_\mathrm{{cc}}^2 +p_\mathrm{{cs}}^2\right).\label{eq:29}
\end{equation} 

\noindent Here $n=1,2,3,\dots $ is the index that counts the different doorway states of the system.
Using Eqs.~(\ref{eq:27}), (\ref{eq:28}) and (\ref{eq:29}) in Eq.~(\ref{eq:26 res gig}), 
one obtains the following approximate expression for ${\mathbb {C}}^{\mathrm{gia}}(\Omega)$:

	\begin{equation}
 	 {\mathbb {C}}^{\mathrm{gia}}_{\mathrm{approx}}\left(\Omega\right) = 
	\left(\frac{\Omega\, h_0\, v_{\mathrm{c}}^2}{\pi r_L^2 E \ell}\right)^2
	\frac{1}{P_s^2+P_c^2},
	\label{eq:30}
	\end{equation}

\noindent which can be rewritten as  

	\begin{equation}
	{\mathbb {C}}_{\mathrm{approx}}^{\mathrm{gia}}\left(\Omega\right)=
	\Lambda^{\mathrm{gia}}\frac{{\beta^{\mathrm{gia}}}^{2}}
	{\left(\Omega-\alpha^{\mathrm{gia}}\right)^{2}+{\beta^{\mathrm{gia}}}^{2}},
	\label{eq:31 lorent gig}
	\end{equation}
where
\begin{equation}
\Lambda^{\mathrm{gia}} =\frac{\Omega^2 h^2 v_\mathrm{c}^2 }{E^2 R^4 \pi^2 \ell^2 \;}\frac{1}{\left(p_{\mathrm{cc}}^2 +p_{\mathrm{cs}}^2 \right)\left(p_{\mathrm{cc}}^2 +p_{\mathrm{cs}}^2 +p_{\mathrm{sc}}^2 +p_{\mathrm{ss}}^2 \right)+\beta^2 }, \label{eq:32} \\
\end{equation}


\begin{equation}
\alpha^{\mathrm{gia}} =\frac{n\pi v_\mathrm{c}}{\ell }- \left(\frac{p_{\mathrm{cc}} /p_{\mathrm{cs}} }{1+\frac{\left(p_{\mathrm{sc}}^2 +p_{\mathrm{ss}}^2 \right)}{\left(p_{\mathrm{cc}}^2 +p_{\mathrm{cs}}^2 \right)}}\right. + \left. \frac{p_{\mathrm{sc}} /p_{\mathrm{ss}} }{1+\frac{\left(p_{\mathrm{cc}}^2 +p_{\mathrm{cs}}^2 \right)}{\left(p_{\mathrm{sc}}^2 +p_{\mathrm{ss}}^2 \right)}}\right),\qquad
\label{eq:33}
\end{equation}

\begin{equation}
{\beta^{\mathrm{gia}}}^2 = \left(\frac{p_{\mathrm{sc}} }{p_{\mathrm{ss}} }-\frac{p_{\mathrm{cc}} }{p_{\mathrm{cs}} }\right)\frac{p_{\mathrm{cc}}^2 +p_{\mathrm{cs}}^2 }{\left(p_{\mathrm{sc}}^2 +p_{\mathrm{ss}}^2 \right)\left(1+\frac{\left(p_{\mathrm{cc}}^2 +p_{\mathrm{cs}}^2 \right)}{\left(p_{\mathrm{sc}}^2 +p_{\mathrm{ss}}^2 \right)}\right)}.
\label{eq:34}
\end{equation}

\noindent Expression~(\ref{eq:31 lorent gig})  is the other expression for the giant resonance we were looking
for. It is similar to that of a Lorentzian function except that here $\alpha^{\mathrm{gia}}$ 
and $\beta^{\mathrm{gia}}$ 
are not constants but complicated functions of the frequency $\Omega$. 
Therefore, ${\mathbb {C}}_{\mathrm {approx}}^{\mathrm{gia}}(\Omega)$
is not an exact Lorentzian function.
It is plotted in Fig.~2 by means of 
magenta points.  
As we can see, its plot is very similar to a Lorentzian.
As it is well known, in a perfect Lorentzian its full width at half maximum 
(FWHM) is equal to $2\beta$. However, in our case we cannot assure {\it a priori} that $2\beta^{\mathrm{gia}}$ is equal to the FWHM of the giant resonance.
In the numerical analysis that we have made 
it was observed that near the frequency of resonance the parameter $\beta^{\mathrm{gia}}$  
varies very  slowly when compared to $(\Omega -\alpha^{\mathrm{gia}})^2$. Consequently 
$\beta^{\mathrm{gia}}$ behaves as a constant. 
So, $2\beta^{\mathrm{gia}}$ evaluated at the frequency of the resonance is approximately equal to the FWHM of the giant resonance. Thus,

\begin{equation}
{\mathrm {FWHM}}^{\mathrm{gia}}\approx 2\beta^{\mathrm{gia}}.
\label{eq:35}
\end{equation}

\noindent  The envelope shown in figure (2) obtained from the analytical expression (26) is indistinguishable from the envelope obtained numerically in reference (19). In addition, it was shown in reference (19) that both, the numerical and experimental calculations, admit the same envelope curve 
with the same ${\mathrm {FWHM}}^{\mathrm{gia}}$, therefore, the above expression reproduces the experimental observation.

From Eqs.~(\ref{eq:18})-(\ref{eq:21}) one concludes that the quotients appearing in 
Eq.~(\ref{eq:34})   
do not depend on $b$. Therefore, $\beta^{\mathrm{gia}}$ is independent of $\lambda$  and 
consequently the strength function
phenomenon is not a dissipative effect.

\subsection{\label{sec:3b}  {\bf Common resonances for a bar with a groove}}

\noindent In this section is derived a simple and compact function  that 
describes each common resonance.
This function will be denoted by ${\mathbb {C}}^{com}(\Omega)$. 
Note that in this case  ${\mathbb {C}}^{com}(\Omega)$ is not an 
envelope as was the case of  ${\mathbb {C}}^{gia}(\Omega)$ of  
the giant resonance, but rather it is the curve itself associated 
with a common resonance of the system.
It should also be noted that we already have the exact function 
(Eq.~(\ref {eq:7 ampl acel})) (or alternatively its approximate form, 
Eq.~(\ref {eq:11 ampl acel aprox})) that describes all the common resonances. 
Then, in principle, one can write that ${\mathbb {C}}^{com}(\Omega)$ is 
equal to the square of $\mathbb {A}$ or of ${\mathbb {A}}_{\mathrm {appox}}$.
Nevertheless, since neither 
(Eq.~(\ref{eq:7 ampl acel}))  nor 
Exp.~(\ref{eq:11 ampl acel aprox}))
explicitly show the value of the width of the resonances and neither 
show the relationship between this width and the coefficient $\lambda$ 
they are not the expressions we want.
Furthermore, these expressions are not as simple or compact as desired. 
What will be done then is to transform
Eq.~(\ref{eq:11 ampl acel aprox})
to obtain  a much simpler but approximately equivalent function.

\vskip3mm

First of all, it should be noted that there are several cause why the bar 
could dissipate energy when it is oscillating in compression. One of them is due to the threads used to hang the bar. To analyze this effect, the threads were placed in different places, including in the nodes
of vibration where, of course, the influence of the supports is minimal. Essentially the same resonance curves were obtained in all cases, so this effect can be neglected.

On the other hand the effect of the air surrounding the bar is not negligible, 
but its effect can be incorporated into our theoretical description
by means of an {\it effective} internal friction coefficient 
as the following two experiments show. In the first one,  
the bar was placed inside a vacuum chamber and as the air was removed the width  
of the resonances was measured. This experiment had to be 
done with a shorter bar than the one 
considered in the rest of our work, due to our limitation 
of not having a vacuum chamber that could hold a bar of length 
$L+\ell+\varepsilon=3. 66\, \mathrm{m}$. 
Therefore a bar of only $0.5\, \mathrm{m}$ was used.  
The results are shown in Fig.~7. The vertical axis shows the 
width of the resonance associated with the compressional mode that has two nodes. 
The horizontal axis shows the pressure in the chamber. 
It can be seen that as the air pressure within the chamber decreases, 
the points tend asymptotically to a horizontal line
which is considerably above the abscissa axis. 
This means that the bar continues to dissipate energy even 
though it finds itself in a vacuum. 
This dissipation must be due to the
internal friction in the bar  and it is clear that this effect will be present regardless 
of the size of the bar. 
So, it is concluded that the width of the resonances 
has at least two origins.  
Therefore, it is to be expected that if the width of a
resonance is calculated from expression~(\ref{eq:7 ampl acel})
with a $\lambda$ value associated only with the internal friction,
the experimental width will not be reproduced.

Figure~8 shows the results of the second experiment in which we return to analyze 
the original bar outside of the vacuum chamber. The figure shows the width values 
of the 18 resonances considered in Figs.~3 and 4.
These widths were calculated for three different values of $\lambda$ as indicated in the inset.
The calculations were done with formula~(\ref{eq:7 ampl acel}).
The corresponding widths measured in the laboratory are also shown (blue circles). 
The value $\lambda = 0.01 \, \mathrm {Pa \cdot s}$ is the one suggested in Ref.~[23].
It is seen that with this value (green stars) the predictions of Eq.~(\ref{eq:7 ampl acel})  
are appreciably smaller than the measured widths.
This difference, as already mentioned, is no surprising and is not necessarily due that  
the value of $\lambda$ suggested in Ref.~[23] is incorrect. But rather because
Eq.~(\ref{eq:7 ampl acel}) was derived considering only internal friction and not the effects of the air. However, Fig.~8 also shows that it is possible to use an effective $\lambda$
that takes into account the two effects simultaneously.
The yellow asterisks show the prediction of Eq.~(\ref{eq:7 ampl acel}) 
using $\lambda =0.12\, \mathrm{Pa}\cdot \mathrm{s}$, which reproduces very well 
the linear fitting (yellow line) of the values measured in the laboratory.
In what follows, the width of the resonances due to a given $\lambda$  
is what we are interested in describing.

	\begin{figure}
	\noindent \begin{centering}
	\includegraphics[width=\textwidth]{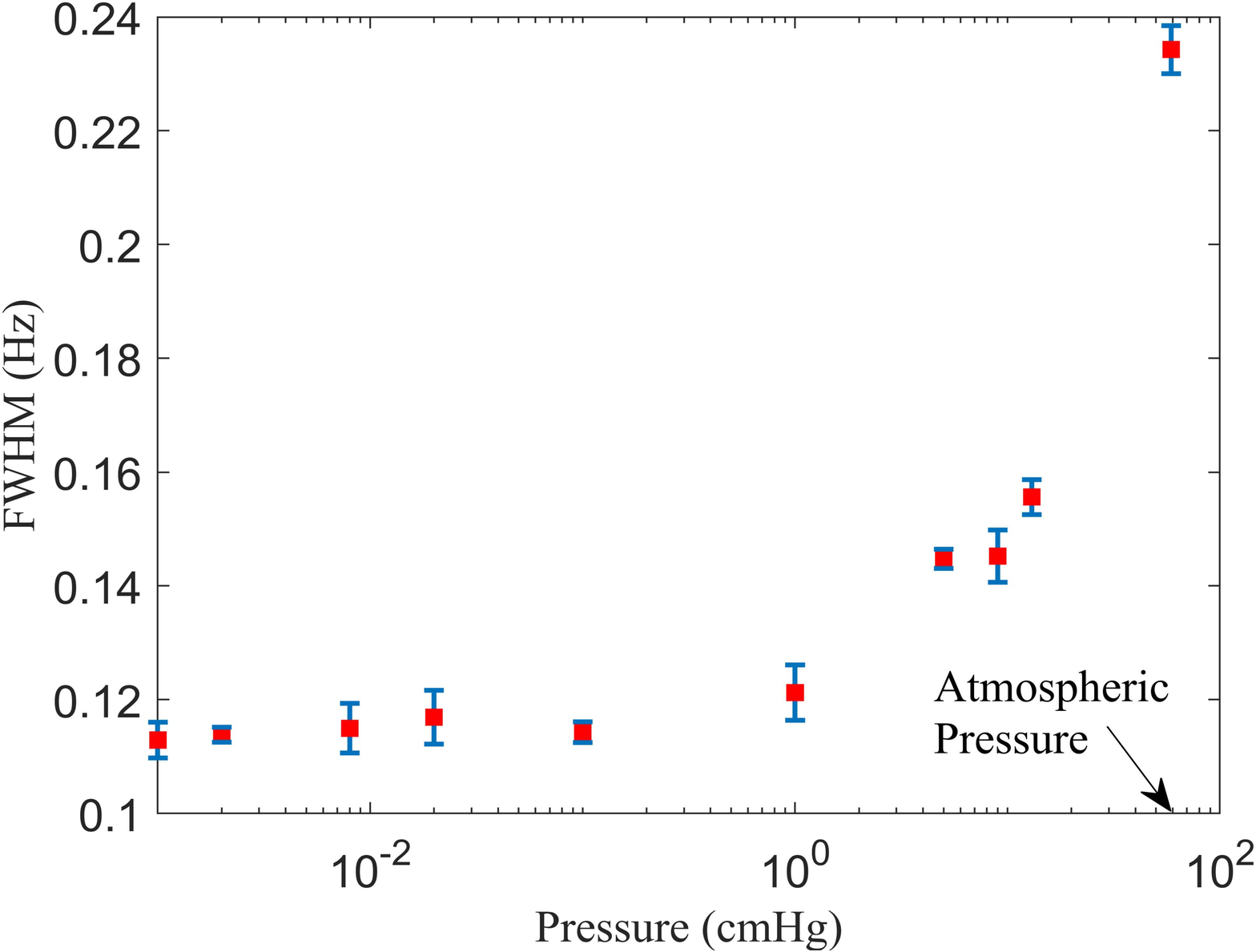}
	\par\end{centering}
	\caption{\label{fig:7}
	Plot of the width of a particular resonance as a function of the pressure 
	inside the vacuum chamber. The analyzed resonance is the one associated 
	with the compressional mode that  has two nodes.
	The width is different from zero even when the pressure is equal to zero.  
	Therefore the bar dissipates energy even 
	if it is  in the  vacuum.}
	\end{figure}

	\begin{figure}
	\noindent \begin{centering}
	\includegraphics[width=\textwidth]{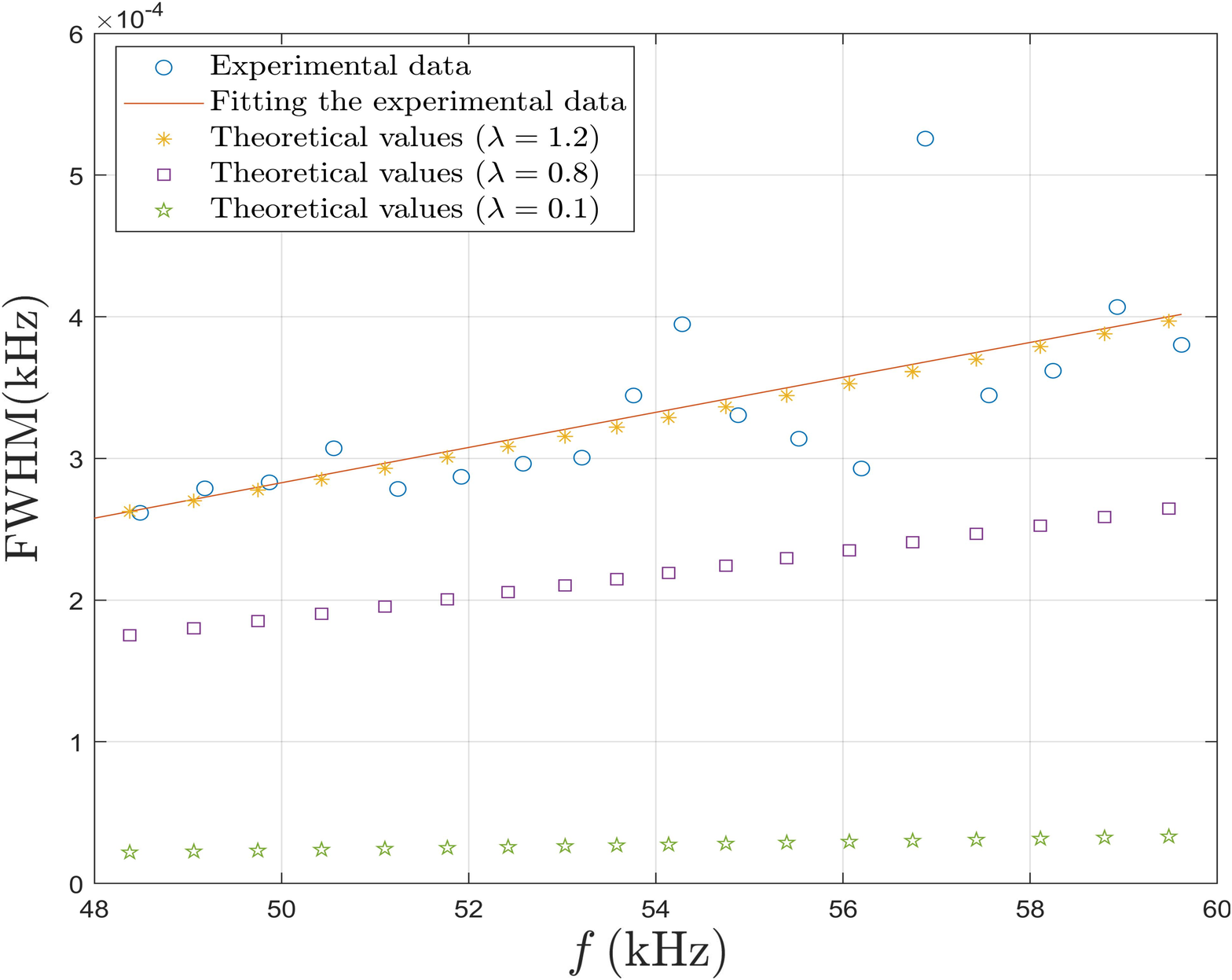}
	\par\end{centering}
	\vskip-4mm
	\caption{\label{fig:8}
	The blue circles indicate the measured widths of the 18 resonances of 
	the large bar with a groove. The horizontal axis shows the value of           
	the frequency. 
	The green stars are the widths of the corresponding  
	18 resonances calculated with Eq.~(\ref{eq:7 ampl acel}) 
	using the value suggested in Ref.~\cite{32}: $\lambda=0.01\,  
	{\mathrm{Pa \cdot s}}$.
	It is seen that with this value the theoretical widths are
	appreciably  
	smaller than the measured widths.
	However, with $\lambda=0.12\,{\mathrm{Pa\cdot s}}$ the calculations 
	(yellow asterisks) reproduce the fitting of the experimental data 
	very well (yellow line). 
    The small squares represent the values corresponding to 
    $\lambda=0.8\, {\mathrm{Pa\cdot s}}
    $.}
	\end{figure}
	
Using Eqs.~(\ref{eq:11 ampl acel aprox}) and (\ref{eq:13}) we get
	\begin{equation}
    {\mathbb {A}}_{\mathrm {approx}}^2=
	\left(\frac{\Omega h_0 v_{\mathrm {c}}}{\pi\, r_L^2\, E}\right)^2
	\frac{1}
	{\left(\mathrm{Re}\,G^{\prime}\right)^{2}
	+\left(\mathrm{Im}\,G^{\prime}\right)^{2}
	},
	\label{eq:36}
	\end{equation}
	\[
	\left(\mathrm{Im}\,G^{\prime}\right)^{2}=
	\sin^{2}\left(\arctan\left(\frac{q_\mathrm{c}}{q_\mathrm{s}}\right)+\frac{\Omega\,L}{v_\mathrm{c}}\right)\,	\left(q_{\mathrm{c}}^2 + q_\mathrm{s}^2\right), 
	\]
where $\left(\mathrm{Re}\,G^{\prime}\right)^{2}$ is given by Eq.~(\ref{eq:12}). 
Now, what matters is not the modulator, but the modulated function. Approximating the function 
$\,\sin^{2}\left(\arctan\left(\frac{q_\mathrm{c}}{q_\mathrm{s}}\right)+\frac{\Omega\,L}{v_\mathrm{c}}\right)\,$ 
by the first term of its Taylor series 
around $\Omega=\frac{v_c}{L}\,\left(m\,\pi-\arctan\left(\frac{q_\mathrm{c}}{q_\mathrm{s}}\right)\right)$, with $m=1~,2~,3~\ldots$, the expression (\ref{eq:36})  reduces to 

\begin{equation}
	{\mathbb {A}}_{\mathrm{approx}}^2\approx  \Lambda^{\mathrm{com}}
       \frac{{\beta ^{\mathrm{com}}}^2 }
       {   
       {\left(\Omega -\alpha^{\mathrm{com}} \right)}^2 + {\beta^{\mathrm{com}}}^2 
       } 
	\equiv {\mathbb {C}}^{\mathrm{com}}(\Omega)
\label{eq:37}
\end{equation}	

\noindent where

\begin{eqnarray}
	\Lambda^{\mathrm{com}} &=& \frac{\Omega^2 \,{h_0 }^2 \,v_c^2 }{{\textrm{E}}^2 \,\left({\mathrm{Re}\,G^{\prime}}			\right)^2 \,r_L^4 \,\pi^2 } \label{eq:38} \\
	{\beta^{\mathrm{com}}}^2 &=& \frac{{\left(\mathrm{Re}\,G^{\prime}\right)^2} \,v_c^2 }{L^2 \,{\left({q_\mathrm{c}}^2 
	+q_{\mathrm{s}}^2 \right)}} \label{eq:39} \\
	\alpha^{\mathrm{com}} &=& 
      \frac{v_c\, \left(\pi \,m-\arctan\left(\frac{q_\mathrm{c}}{q_\mathrm{s}}\right)\right)}
      {L}  \label{eq:40}
\end{eqnarray}

\noindent Equation~(\ref{eq:37}) is the expression for the common resonance we were looking
for. As was the case of the giant resonance, it is similar to that of a Lorentzian 
function except that $\alpha^{\mathrm{com}}$ and $\beta^{\mathrm{com}}$ are not constants but
complicated functions of the frequency $\Omega$. 
The red points of Figs.~3 and 4 are plots of ${\mathbb {C}}^{\mathrm{com}}(\Omega)$ 
corresponding to the 18 common resonances of Fig.~2. 
As we can see, although ${\mathbb {C}}^{\mathrm{com}}(\Omega)$  
is not an exact Lorentzian function, their plots (red points) are very similar to it.
Furthermore, when they are compared with 
$\mathbb {A}^2$ given by  Eq.~(\ref{eq:7 ampl acel}) (blue continuous lines); an excellent agreement is evident. Using the same arguments that led from equation (34) to (35) we can establish a similar result for the common resonances, that is, 

\begin{equation}
\mathrm {FWHM}^{\mathrm{com}}\approx 2\beta^{\mathrm{com}}.	
\label{eq:41}
\end{equation}	

\noindent From Eqs.~(\ref{eq:12}), (\ref{eq:14}), (\ref{eq:15}), (\ref{eq:18})-(\ref{eq:21}) it can be 
seen that $\mathrm {Re} G'$  depends linearly on $b$. Therefore,  from Eqs.~(\ref{eq:3}) and 
(\ref{eq:39}), 
it follows that $\beta^{\mathrm{com}}$ is a linear function of      
$\lambda$. Consequently the width of the common resonances 
predicted by this model is due to a dissipative effect. 
Thus, our formulation explicitly reproduces the following two experimental observations:

\noindent (1) The width of common resonances depends on the value of $\lambda$

\noindent (2) The width of the giant resonances is independent of $\lambda$.

Thus, the giant resonance is unaffected to changes in the width of common resonances. It is also unaffected to changes in the atmospheric pressure surrounding the bar. The latter can be concluded by looking at figure (8), which shows that by changing the pressure, there is a change in the width of the common resonances, but this, as already said, does not affect the giant resonance.

Furthermore, numerically analysing the behaviour of the function defined in expression (7) it has been observed that the length of the longest bar has a negligible effect on the giant resonances (except when the value of $\eta$ is very small as compared to the value we used). Figure 9 shows that varying the length $L$ changes the separation between the common resonances but the envelope in all four figures is practically the same. 
The small differences are only noticeable when doing a more detailed analysis of these envelopes. In the language of fuzzy structure theory we can say that the details of the fuzzy couplings (in this case the length of the longest bar, the internal friction coefficient and the atmospheric pressure) do not have an important influence on the global or macroscopic behaviour of the system. In  contrast, the changes in the length of the shortest bar ({\it i.e.} the changes in the master structure) do affect the giant resonance. It has been observed, in fact,  that by changing the length $\ell$ the giant resonances change its position.

\begin{figure}
    \includegraphics[width=\textwidth]{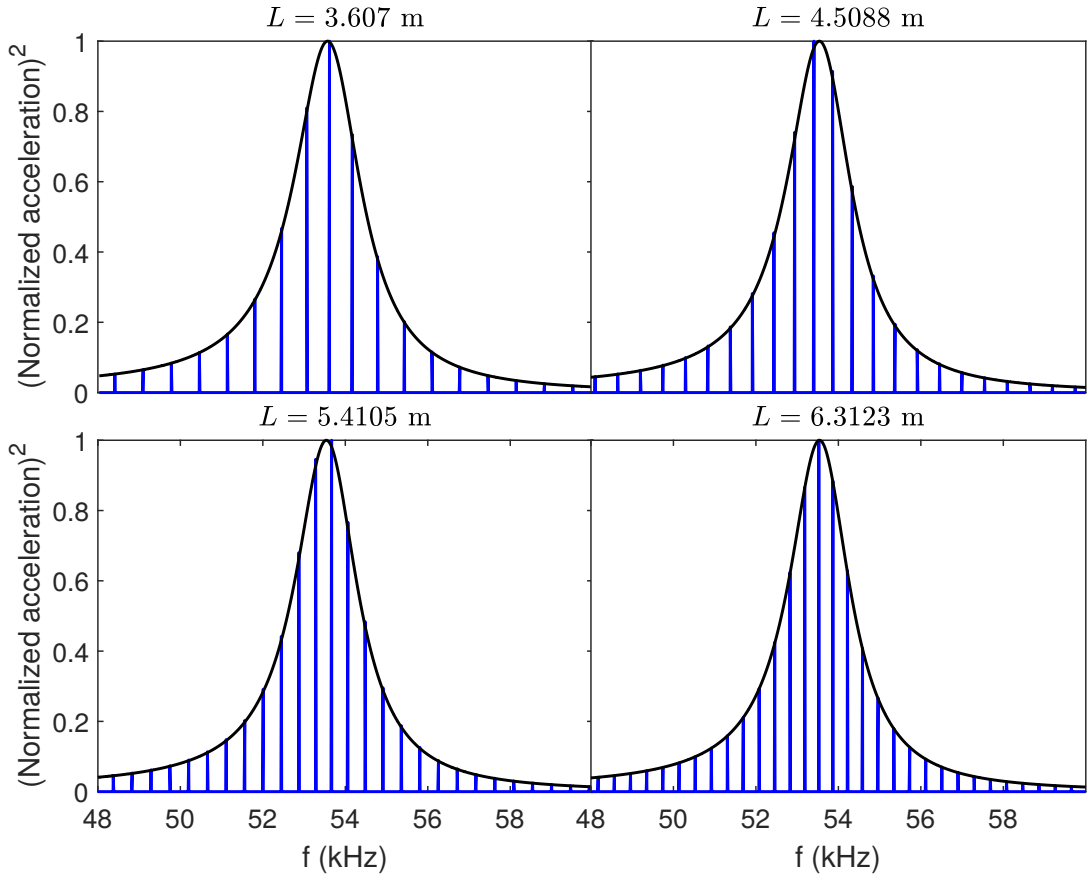}
	\caption{\label{fig:9}
	Four plots of the common resonances and their envelope curves (giant resonances) for the bar of
	Fig.~1 for different values of the largest bar of length $L$. If $L$ changes the same giant resonance
	is obtained. It only changes the separation among the common resonances.}
	\end{figure}

In the previous discussion we have considered a bar composed of three coupled cylinders because the objective was to analyse the experimental observations and the theoretical study that were made on the system discussed in reference [8]. But giant resonances, doorway states and the strength function phenomenon also exist in simpler bars consisting of only two coupled cylinders. In fact, if one bar of Fig.~(1) is removed, for example, the central bar that joins the two end bars of lengths $3.607\, {\rm m}$ and $0.0498\,{\rm m}$ and it is make the first narrower, the bar shown in Fig.~(10) is obtained. Then, when calculating the response of this new system one obtains Fig.~(11a). The vertical axis, as before, shows the values of the square amplitude of the acceleration at the left end of the system as a function of the frequency when the system is excited by the force given in Eq.~(1). The presence of three giant resonances is observed.
For the calculation the equation (7) was used, taking one of the length equal to zero to adapt it to the case of only two cylinders. 

We have also used our formulation to compare the predictions of the fuzzy structure theory with the exact calculation. For this, we have considered again the bar of Fig.~(10) but now with the values used in 
Ref.~[nunes]. The results are shown in Figs.~11(b) and 11(c). The presence of two giant resonances is observed. Fig.~(11c) should be compared with figures (4) of reference [nunes] where the results of the fuzzy structure theory are presented. Both studies show that the response of the system  is appreciably higher around the frequencies $39\,{\rm kHz}$ and $75\,{\rm kHz}$. This comparison convinces us of the usefulness of the fuzzy structure theory and confirms its validity. Nevertheless, as expected, the details are very different. But that should not cause any concern because, from the beginning, the fuzzy structure theory establishes that its objective is to obtain global results leaving aside the details. Therefore the comparison is very satisfactory.

\begin{figure}
	\noindent \begin{centering}
	\includegraphics[scale=.6]{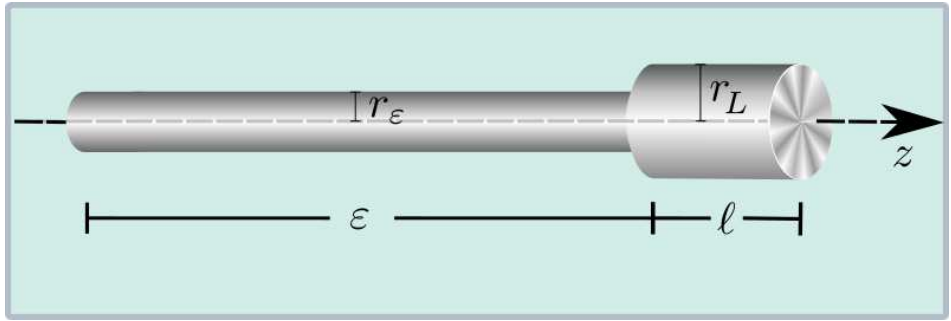}
	\par\end{centering}
	\caption{\label{fig:10}
	The bar is formed by two circular cylinders as the one considered in figure 2 
	of Ref.~[numes].}
	\end{figure}

\begin{figure}
	\noindent \begin{centering}
	\includegraphics[width=\textwidth]{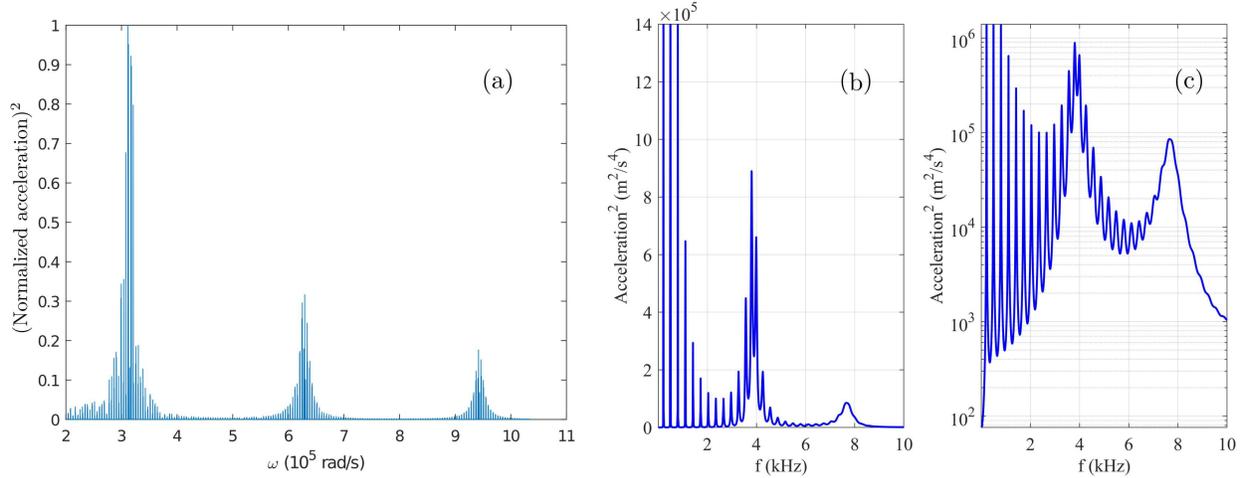}
	\par\end{centering}
	\vskip-2mm
	\caption{\label{fig:11}
	These figures show the values of the square amplitude of the acceleration at the left end
    of the bar of Fig.~(10). The vertical axis scale in Fig.~(11c) is logarithmic. 
    The lengths of the cylinders corresponding to Fig.~(11a) are $0.0492\,{\rm m}$ and $3.607\,{\rm m}$.
    While for the case of Figs.~(11b) and (11c) are $0.2\,{\rm m}$ and $2.46\,{\rm m}$.
    Three giant resonances are observed in Fig.~(11a) and two in Figs.~(11b) and (11c).
    The latter is the case studied in the reference [numes]}.
	\end{figure}
	
\vskip6mm

\subsection{\label{sec:3c} {\bf Common resonance curves and their envelope curve for a bar without a groove}}

\vskip4mm

Because a bar without a groove can be considered as a particular case of a bar 
with a groove when $\eta=1$, we can use the expressions obtained 
previously and take $\eta=1$. Then the expression for 
the acceleration at the left end of the bar 
is again given by  Eqs.~(\ref{eq:6}) and (\ref{eq:7 ampl acel}). 	   
But in this case the function $G$ is much simpler. 
By doing $\eta=1$ in Eq.~(\ref{eq:5}) we get $c_1=1$ and $c_2=c_3=c_4=0$. Therefore, 

	\begin{equation}	
	G(x(i\Omega))=\sinh \left(\frac {i\Omega\mathcal {L}}
	{v_c  \sqrt{ 1+ib  }}
	\right ),
	\label{eq:42}
	\end{equation}

\noindent where ${\mathcal{L}}=L+\ell +\epsilon$ is the total length of the bar without a groove. 
Its radius will be denoted by $R$. Then Eq.~(\ref{eq:42}) can be written as 

	\begin{equation}	
	G(x(i\Omega))=\sinh \left(\frac {i\Omega\mathcal {L}}
	{v_c  \sqrt{ 
	\sqrt{1+b^2}\exp(i\,\,{\arctan{b})} 
	}}\right )
	\label{eq:43}
	\end{equation}

	\begin{equation}
       =\sinh \left(\frac {\Omega\mathcal {L}\left[
	   \sin{   \{\frac{1}{2} \arctan{b}\} }
      +i\cos{  \{\frac{1}{2} \arctan{b}\} }\right]
      }
	{v_c   
	(1+b^2)^{1/4} } 
	\right ).
	\label{eq: 44}
	\end{equation}

\noindent Using the equalities 

	\[ 
	 \sin   \left(\frac{1}{2} \arctan{b}\right)=\frac{1}{\sqrt 2}\frac{1}{(1+b^2)^{1/4}}\sqrt{ \sqrt{1+b^2}-1  };
	 \]
	\[
	\cos   \left (\frac{1}{2} \arctan{b}\right)=\frac{b}{\sqrt 2}\frac{1}{(1+b^2)^{1/4}}\frac{1}{\sqrt{ \sqrt{1+b^2}-1  }},
	\]
       
\noindent we obtain 

	\[
	\mathrm{Re}\,G
	=\sinh\left(  \frac{1}{\sqrt 2}~
      \frac{\Omega \mathcal {L} \sqrt{  \sqrt{1+b^2}-1   }}{v_{\mathrm {c}} \sqrt{1+b^2}}
	\right)\times
	\]
	\begin{equation}
	\hskip2cm
	\cos\left(  \frac{b}{\sqrt 2}~
      \frac{\Omega \mathcal {L}}{v_{\mathrm {c}} \sqrt{1+b^2}}\frac{1}
	{\sqrt{  \sqrt{1+b^2}-1   }}\right),
	\label{eq:45}
	\end{equation}
	
	\[
	\mathrm{Im}\,G
	=\sin\left(  \frac{b}{\sqrt 2}~
      \frac{\Omega \mathcal {L}}{v_{\mathrm {c}} \sqrt{1+b^2}}\frac{1}
	{\sqrt{  \sqrt{1+b^2}-1   }}\right)\times
	\]
	\begin{equation}
	\hskip2cm
	\cosh\left(  \frac{1}{\sqrt 2}~
      \frac{\Omega \mathcal{L}}{v_{\mathrm {c}} \sqrt{1+b^2}}
	\sqrt{  \sqrt{1+b^2}-1   }\right).
	\label{eq:46}
	\end{equation}

\noindent Since $b=\lambda\,\Omega/E \ll 1$ we use the following approximations 

	\[
	(1+b^2)^{1/2}\approx 1+b^2/2; \quad (1+b^2)^{1/4}\approx 1+b^2/4; 
	\]
	\[
	\quad (1+b^2)^{-1/2}\approx 1-b^2/2,
	\]

\noindent which implies 

	\[
	\frac{\sqrt{\sqrt{1+b^2}-1}}{\sqrt{1+b^2}}\approx \frac{b}{\sqrt 2};\qquad 
	\frac{b}{\sqrt{1+b^2}}\frac{1}
	{\sqrt{\sqrt{1+b^2}-1}}\approx \sqrt{2}.
	\]

\noindent The substitution of this into Eqs.~(\ref{eq:45}) and (\ref{eq:46}) gives

	\[
	\mathrm {Re}\, G\approx \sinh \left( \frac{\Omega{\mathcal{L}}    b}{2v_c}  \right) 
    \cos  \left(  \frac{\Omega{\mathcal L}   }{v_c}  \right) \approx  \frac{\Omega{\mathcal L}    b}{2v_c} \cos  \left(  \frac{\Omega{\mathcal L}   }{v_c}  \right)
   \]

	\[
	=\frac{\Omega^2{\mathcal L} \lambda}{2v_cE} \cos  \left(  \frac{\Omega{\mathcal L}   }{v_c}  \right),
	\]

	\[
	\mathrm {Im}\, G\approx\cosh \left( \frac{\Omega{\mathcal L}    b}{2v_c}  \right) 
                                \sin  \left(  \frac{\Omega{\mathcal L}   }{v_c}  \right)\approx
	                          \sin  \left(  \frac{\Omega{\mathcal L}   }{v_c}  \right).
	\]
	
\noindent With these values the expression for the acceleration amplitude Eq.~(\ref{eq:11 ampl acel aprox}) (denoted as ${\mathbb {A}}_{\mathrm{approx}}^{\prime}$) becomes  

	\begin{equation}
	{\mathbb {A}}_{\mathrm{approx}}^{\prime}=
	\frac{\Omega h_0 v_c}{\pi R^2\,E} 
	\frac{1}{\sqrt{ \left(\frac{\Omega^2\mathcal{L} \lambda}{2v_c E}\right)^2  
	\cos^2\left(\frac{\Omega\mathcal{L}}{v_c}\right) +  
	\sin^2\left(\frac{\Omega\mathcal {L}}{v_c}\right)    }},
	\label{eq: 47}
	\end{equation}
	
This is another of the expressions that we wanted to obtain. 
We see that the relative maxima of ${\mathbb {A}}_{\mathrm{approx}}^{\prime}$ 
occur when $\cos^2\left(\frac{\Omega\mathcal{L}}{v_c}\right)=0$ and when
                 $\sin^2\left(\frac{\Omega\mathcal{L}}{v_c}\right)=0$. 
However, since the value of the factor 

  	\begin{equation}
	\left(\frac{\Omega^2\mathcal{L} \lambda}{2v_c E}\right)^2  
	\end{equation}

\noindent is negligible as compared to 1 (it is on the order of $10^{-10}$), 
it follows that the maxima of ${\mathbb {A}}_{\mathrm{approx}}^{\prime}$ due to the zeros of $\cos^2\left(\frac{\Omega\mathcal{L}}{v_c}\right)$
are negligible as compared to the maxima due to the zeros of 
$\sin^2\left(\frac{\Omega\mathcal{L}}{v_c}\right)$. Therefore,  
significant maxima of ${\mathbb {A}}_{\mathrm{approx}}^{\prime}$
occur only when $\sin^2\left(\frac{\Omega\mathcal{L}}{v_c}\right)=0$.  
These values occur with a frequency $\pi v_c/{\mathcal {L}}$ which is equal to the separation
between the dashed blue vertical lines in Fig.~5). In these maxima the value of 
${\mathbb {A}}_{\mathrm{approx}}^{\prime}$ is 

	\begin{equation}
	\frac{2 h_0 v^2_c}{\pi R^2\,\lambda \mathcal{L} }~ \frac{1}{\Omega}
	\end{equation}

\noindent and therefore the equation for the envelope that passes through these maxima is 

	\begin{equation}
	{\mathbb{C}}^{\prime}(\Omega)=\left(\frac{2 h_0 v^2_c}{\pi R^2\,\lambda \mathcal {L} }\right)^2
	 \frac{1}{\Omega^2}.
	\label{eq:50}
	\end{equation}

\noindent This function is plotted in Fig.~5 as a black line. It is clearly shown that this line passes through the maximum values of each resonance. Since for this bar the envelope is described by an expression that
depends on the inverse square of the frequency, the strength function phenomenon is not present.
Note that eliminating the fast oscillations from Eq.~(\ref{eq: 47})  by doing 
$\cos^2(\Omega\mathcal{L}/v_{\mathrm {c}})=1$ 
and 
$\sin^2(\Omega\mathcal{L}/v_{\mathrm{c}})=0$ in order to obtain the 
envelope, is similar to what was done  
for the general case $\eta\neq 1$ where the giant resonance was obtained. 

It is worth noting that the envelope shape for the case of bars without groove has an important influence on the envelope shape for the case of bars with a groove. Indeed,  the decreasing shape of the black curve in Fig. 5 is responsible for the asymmetry of the black  curve of Fig.~2 (the left tail is higher than the right tail). 

\vskip3mm

We will now derive an approximate and compact expression for Eq.~(\ref{eq: 47})
that explicitly shows its similarity to a Lorentzian function.
To do this, we analyse the behavior of expression (\ref{eq: 47})  in the vicinity of the natural frequencies of the bar. 
These are: $\Omega_n=2\pi f_n= \pi v_{\mathrm {c}}\, n /{\mathcal {L}}$. 
Therefore, only values of $\Omega$ within the interval 
	\begin{equation}
	I_n=[\Omega_n-\Delta, \Omega_n+\Delta],
	\end{equation}

\noindent will be considered, being $\Delta$ a small but appropriate frequency interval.
So $\Omega=\Omega_n\pm \delta$, with $0\leq \delta \leq \Delta$. Then, 
	\begin{equation}
	\cos^2 \left(\frac{{\mathcal {L}}\Omega}{v_{\mathrm{c}}}\right)\approx 
	\cos^2 \left(\frac{{\mathcal {L}}\Omega_n}{v_{\mathrm{c}}}\right)=
	\cos^2 \left(\pi n\right)=1
	\end{equation}
and
	\[
	\sin^2 \left(\frac{{\mathcal {L}} \Omega              }{v_{\mathrm{c}}}\right)= 
	\sin^2 \left(\frac{{\mathcal {L}}(\Omega_n\pm\delta)}{v_{\mathrm{c}}}\right)=
	\sin^2 \left(\frac{\pm {\mathcal {L}} \delta }{v_{\mathrm{c}}}\right)
	\]
	\begin{equation}
	\approx \left(\frac{{\mathcal {L}} \delta }{v_{\mathrm{c}}}\right)^2=
	\left(\frac{{\mathcal {L}}\Omega}{v_{\mathrm{c}}} -n\pi \right)^2
	\end{equation}
 
\noindent Substituting these equalities in Eq.~(\ref{eq: 47}) one obtains 

	\begin{equation}
	{\mathbb {A}}_{\mathrm{approx}}^{\prime}\approx
	\frac{\Omega h_0 v_c}{\pi R^2\,E} 
	\frac{1}
	{\sqrt{ \left(\frac{\Omega^2\mathcal{L} \lambda}{2v_c E}\right)^2  
	+  \left(\frac{{\mathcal {L}}\Omega}{v_{\mathrm{c}}} -n\pi \right)^2 }}
	\label{eq: 54}
	\end{equation}

\noindent and

	\begin{equation}
	{\mathbb {A}}_{\mathrm{approx}}^{\prime 2}\approx
	\left(\frac{2  h_0 v^2_c}{\pi R^2\,\lambda {\mathcal{L}}\Omega}\right)^2 
	\frac{\beta^{\prime 2}}
	{  (\Omega-\Omega_n )^2+\beta^{\prime 2 }  },
	\label{eq: 55}
	\end{equation}

\noindent where

	$$\beta^{\prime }=\frac{\lambda \Omega^2}{2E}.$$

\noindent Eq.~(\ref{eq: 55}) is the approximate expression for the curve associated with the common 
resonance centered on $\Omega_n$  and it is another of the expressions 
that we wanted to obtain. The plot of expression (\ref{eq: 55}) reproduces each of the blue 
lines in Fig.~ 5 very well. These lines were calculated with the exact formula
(\ref{eq:7 ampl acel}) with $\eta=1$. As was the case of the 
other expressions for the resonant curves,  expression (\ref{eq: 55}) is not an exact Lorentzian function because $\beta^{\prime }$ is not a constant. However, 
using again the same arguments that led from equation (34) to (35) we can establish 
the following result for the full width at half maximum,  
denoted as ${\mathrm{FWHM}}^{\prime }$, of the common resonances for a bar without a groove
	\begin{equation}
	{\mathrm{FWHM}}^{\prime}\approx 2\beta^{\prime }=\frac{\lambda \Omega^2_n}{E}.
	\label{eq: 56}
	\end{equation}

\noindent This result explicitly shows that the coefficient $\lambda$ determines the width of the common resonances. This, of course, was to be expected based on the previous discussion of the bars with a groove.

\vskip5mm

\section{\label{sec:7} CONCLUSIONS}

Analytical expressions for the different resonances present in vibrating elastic systems consisting of  coupled rods, have been derived. The most important of these expressions is the one associated with the giant resonances, that is, the analytical expression for the envelope of the common resonances, which has not been previously discussed in the literature. It was shown that in these systems one of rod provides the doorway state and the others the sea of states within which the external excitations are distributed, giving rise to a giant resonance. This same situation, contemplated from the point of view of the fuzzy structure theory, shows that in the system of coupled rods one of them acts as the master structure and the others as fuzzy structures. Nevertheless, in our case the exact expressions obtained allow us to verify that the approximate predictions of the fuzzy theory are reasonable. 

Closed expressions for the $\mathrm{FWHM}'s$ of the resonance curves were also derived. The phenomenon of the strength function was also analysed.
In the case of common resonances, the internal friction coefficient $\lambda$ of the Voigt model explicitly appears in the expression for $\mathrm{FWHM}^{\mathrm{com}}$, meaning that the width of the common resonances is due to the energy dissipation.  
On the other hand, for the case of the giant resonance, the internal
friction coefficient does not appear in the expression for $\mathrm{FWHM}^{\mathrm{gia}}$, meaning that the strength function phenomenon is not a dissipative effect.

The effect of fuzzy couplings is to distribute the excitations applied to the master structure, among the states of the composite system, in such a way that the intensity with which they are excited has an envelope with a quasi-Lorentzian shape. The width of this quasi-Lorentzian is independent (surprisingly) of the dissipation factor of the master structure.

If one attempts to measure a microscopic property with a macroscopic meter (with low resolution), the conditions are met for a giant resonance to exist and that is what the meter will detect.

The formulation derived here is the continuation of a previous work~\cite{28} in which the giant resonance was also discussed but without having an analytical expression to describe it.  In this work, this concept is discussed from analytical, numerical  and experimental perspectives.
In conclusion, for the elastic systems case with a doorway state, giant resonances are fully understood.

\vskip5mm
\section*{Author's Contributions}
All authors contributed equally to this work
\section*{acknowledgments}
The authors want to thank the project  DGAPA-PAPIIT IN111019.\\
J.A.R. thanks CONACYT-Mexico for the fellowship for doctoral studies.\\
E.A.C. acknowledges CONACYT-Mexico for the postdoctoral research fellowship at IFUNAM.


\section{APPENDIX}
This appendix briefly reproduces the derivation 
of the formulas governing the compressional oscillations in a bar with a groove 
developed in Ref.~\cite{28}
The starting point is Newton's second law applied to the description of compressional 
oscillations in a circular cylinder whose cross section area is equal to $a$~\cite{32}:

	\begin{equation}
	\frac{\partial \sigma}{\partial z}-\rho\frac{\partial^2 A(z,t)}{\partial t^2}
	=-\frac{ F(z,t)}{a}
	\label{eq:a1}
	\end{equation}

\noindent where $\sigma$ is the  stress, $A(z,t)$ the displacement in time $t$ (suffered by the portion of material originally located at point $z$ on the axis $Z$) 
when the bar is oscillating, $S=\partial A/\partial z$  the strain and $F(z,t)/a$
the external force applied per unit volume.
The speed of the deformation is 
$\partial A(z,t)/\partial t$. The speed of propagation of the waves will be denoted as 
$v_{c}$.
Because the excitation acts continuously, the dissipation of
energy plays an important role in the experiment by preventing
an unlimited growth of the response. In the
analytical description of the phenomenon this effect was included by means of a model. 
In the literature there are several models.
In the formulation of Ref.~\cite{28} the Voigt viscoelastic model was used.
It consists of assuming that the total stress applied $\sigma$ is the sum of the 
stress associated with deformation plus a stress associated with viscosity. Thus, the constitutive relation is \cite{31,32,33}

	\begin{equation}
	\sigma=ES+\lambda\frac{\partial S}{\partial t}
	\label{eq:a2}
	\end{equation}

\noindent here $\lambda$ is the coefficient of viscosity.
Substituting the definition of $S$ in Eq.~(\ref{eq:a2}) we have

	\begin{equation}
	\sigma =E\frac{\partial A(z,t)}{\partial z}+
	\lambda\frac{\partial}{\partial z}\frac{\partial A(z,t)}{\partial t}
	\label{eq:a3}
	\end{equation}

\indent Then, the equation of motion that governs the
compressional vibrations in each cylinder is

	\begin{equation}
	\frac{\partial^2 A(z,t)}{\partial z^2}+\frac{\lambda}{E}\frac{\partial}{\partial t}
	\frac{\partial^2A(z,t)}{\partial z^2}-\frac{\rho}{E}\frac{\partial^2A(z,t)}{\partial t^2}=-\frac {F(z,t)}{Ea}.
\label{eq:a4}
	\end{equation}

\noindent It is easy to see that when the bar is excited at its right extreme 
$z=L+\epsilon+\ell~$ by the 
force given by Eq.~(1)  the function $F(z,t)$ 
of the above equation must be equal to $F(z,t)=h_0\sin(\Omega t) \delta(x-L-\epsilon-\ell)\Theta(t)$.
\noindent If we take the Laplace transform of Eq.~(\ref{eq:a4}) for each cylinder, then solve the three resulting equations and apply the boundary conditions, we obtain the Laplace transform
for the function $A(z,t)$ for the full bar:

	\begin{equation}
	{\mathscr {A}}(z,s)={\mathscr L}\{A(z,t)\}(s).
	\label{eq:a5}
	\end{equation}

\noindent Finally, the expression for  the acceleration 
$\partial^2 A(z,t)/\partial t^2$  
was obtained 
using the Mellin inversion integral with the Bromwich 
contour. The expression turned out
quite complicated (see Eq.~(21) of Ref. \cite{28}), but if one ignores the transitory part, the following expression for the acceleration at the left end of the bar is obtained; which, although still complicated, is more manageable,

	\begin{equation}
	\frac{d^2A(0,t)}{dt^2}=\frac{v_c h_0}{a_LE}\Omega\mathrm{Re} 
	\left\{ \frac {\sqrt {1+\frac{i\lambda \Omega}{E}}e^{i\Omega t}}{G(x(i\Omega))}\right\}
	\end{equation}

\noindent where

	\begin{equation}
	x(i\Omega)=\frac{i\Omega}{   \sqrt{1+\frac{i\lambda \Omega}{E}}     },
	\end{equation}

	\begin{equation}
	v_c=\sqrt{\frac{E}{\rho}}
	\end{equation}
                                                          
\noindent and

	\begin{eqnarray}
	G(Y)=&\frac{1}{4}\left(\eta^2+\frac{1}{\eta^2}+2\right)\sinh\frac{Y(L+\ell+\epsilon)}{v_c}&
	\nonumber\\	
	&-\frac{1}{4}\left( \eta^2+\frac{1}{\eta^2} -2\right)\sinh\frac{Y(L+\ell -\epsilon)}{v_c}& \nonumber\\	
	&+\frac{1}{4}\left( \eta^2 -\frac{1}{\eta^2} \right) \left( \sinh\frac{Y(L-\ell+\epsilon)}{v_c}\right. &\nonumber\\	
	&\qquad-\left. \sinh\frac{Y(L-\ell-\epsilon}{v_c}\right)&
	\end{eqnarray}

	\begin{eqnarray}	
	&=&\left[\frac{1}{\eta^2}\sinh\frac {YL}{v_c}\sinh\frac{Y\ell}{v_c} 
                            +  \eta^2 \cosh\frac{YL}{v_c}\cosh\frac{Y\ell}{v_c}\right]\sinh\frac{Y\epsilon}{v_c}\nonumber\\ 
      &&+\sinh\frac{Y[L+\ell]}{v_c}\cosh\frac{Y\epsilon}{v_c}	
	\end{eqnarray}	

\noindent being $a_L=\pi r_L^2$ the cross-section area of the cylinders 1 and 3.      

\nocite{*}
\bibliography{sampbib}

\end{document}